\RequirePackage{fix-cm}

\documentclass[times]{article}

\usepackage{graphicx}
\usepackage{stmaryrd}

\usepackage{amssymb}
\usepackage{amsmath}
\usepackage{eucal}

\usepackage[ruled,vlined]{algorithm2e}
\usepackage[driverfallback=dvipdfm,breaklinks=true,pdfcreator={LaTeX e Dvips}]{hyperref}

\usepackage{float}

\title{Ambiguous phase assignment of discretized 3D geometries in topology optimization}

\author{Jorge L. Barrera, Kurt Maute \\[12pt]
Ann and H. J. Smead Department of Aerospace Engineering Sciences,\\ 
University of Colorado at Boulder, 3775 Discovery Dr, Boulder, CO 80303, USA\\
Corresponding author: maute@colorado.edu}

\begin{document}

\maketitle

\begin{abstract}
Level set-based immersed boundary techniques operate on nonconforming meshes while providing a crisp definition of interface and external boundaries. 
In such techniques, an isocontour of a level set field interpolated from nodal level set values defines a problem's geometry.
If the interface is explicitly tracked, the intersected elements are typically divided into sub-elements to which a phase needs to be assigned.
Due to loss of information in the discretization of the level set field, certain geometrical configurations allow for ambiguous phase assignment of sub-elements, and thus ambiguous definition of the interface.
The study presented here focuses on analyzing these topological ambiguities in embedded geometries constructed from discretized level set fields on hexahedral meshes.
The analysis is performed on three-dimensional problems where several intersection configurations can significantly affect the problem's topology. This is in contrast to two-dimensional problems where ambiguous topological features exist only in one intersection configuration and identifying and resolving them is straightforward.
A set of rules that resolve these ambiguities for two-phase problems is proposed, and algorithms for their implementations are provided.
The influence of these rules on the evolution of the geometry in the optimization process is investigated with linear elastic topology optimization problems.
These problems are solved by an explicit level set topology optimization framework that uses the extended finite element method to predict physical responses.
This study shows that the choice of a rule to resolve topological features can result in drastically different final geometries.
However, for the problems studied in this paper, the performances of the optimized design do not differ.
\end{abstract}

%% \linenumbers

%-------------------------------------------------------------------------------------------------%
\section{Introduction} \label{sec:intro}
%-------------------------------------------------------------------------------------------------%

In topology optimization, a parametrized field that describes a problem's geometry can freely evolve in search of an optimal design.
The level set method (\cite{osher1988fronts,sethian1996level,van2013level}) provides a convenient framework for this process.
Traditionally, immersed analysis techniques smear the interface over elements (e.g., the Ersatz material method \cite{allaire2005structural}). 
However, a rapidly growing class of these techniques are able to capture and track the interface explicitly (e.g., cut finite element method \cite{burman2015cutfem}, extended finite element method \cite{belytschko2009review}). 
In such techniques, embedding the geometry in a computational domain via a discretized level set field ($\text{LSF}$) may allow for ambiguous definitions of the discretized geometry at any stage of the design optimization process.
Therefore, methods for detecting and resolving these ambiguities are needed.
Yet, with only a few exceptions (\cite{Allaire2013,Nguyen2018,laurain2018analyzing}), work on level set topology optimization omits mentioning and addressing this issue.

\begin{figure}[ht!]
	\centering
	\includegraphics[width=1.0\linewidth]{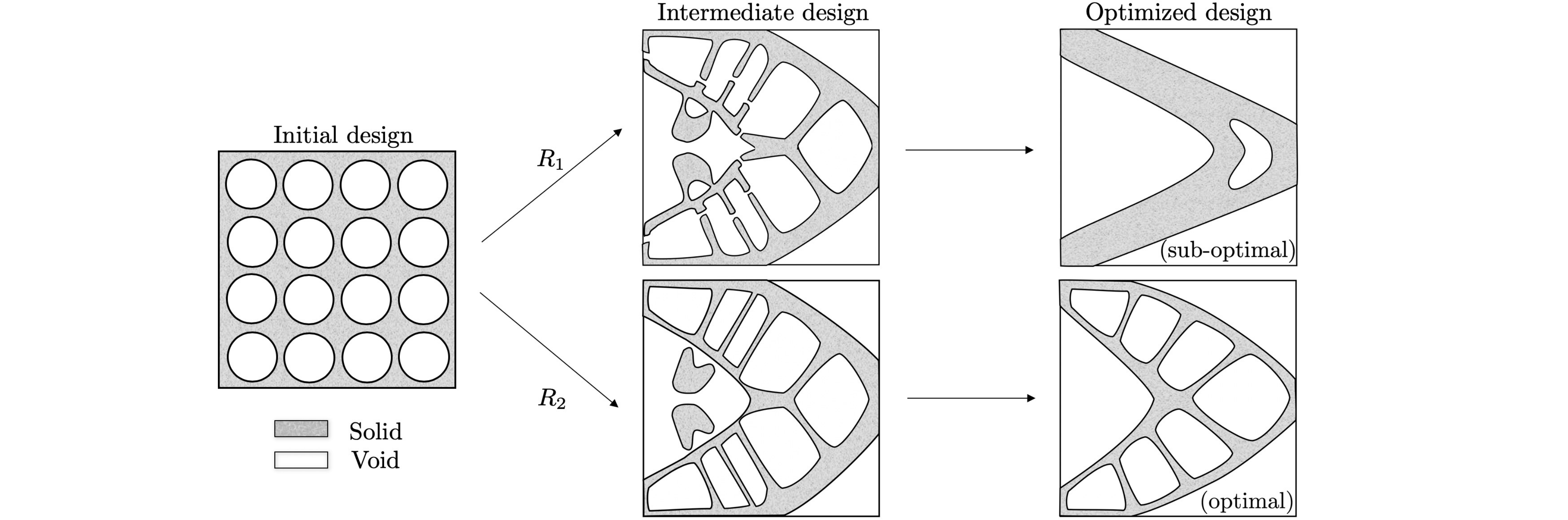}
	\caption{Probable effect of phase assignment using either rule $R_1$ or $R_2$ to resolve ambiguous geometrical features in the optimization process.}
	\label{fig:sketchForQuestionThisPaperAnswers}
\end{figure}

Ambiguities of discretized geometries can be resolved via user-defined rules.
The effect of a given rule on the optimization process and the optimized designs, i.e. whether some user-defined rules can lead to suboptimal designs, is still an open question.
This scenario is illustrated in the 2D schematics of Fig.~\ref{fig:sketchForQuestionThisPaperAnswers}, where two different rules, $R_1$ and $R_2$, are used to resolve geometric ambiguities.
The sketch shows that, starting from the same initial design, not only different intermediate designs are possible, but also that the optimized designs may differ in geometry and performance.
Hence, in this work we investigate the influence of these rules used to determine ambiguous geometrical configurations in the context of topology optimization.

%Figure

\begin{figure}[b]
	\centering
	\includegraphics[width=1.0\linewidth]{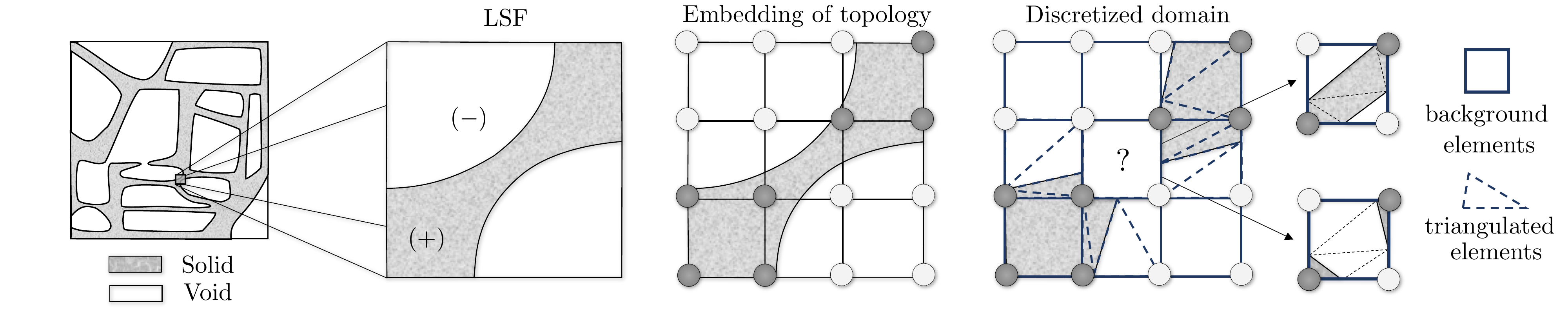}
	\caption{Topology consistency issue in 2D.}
	\label{fig:elemAmbIssue2D}
\end{figure}

In this paper, we focus on level set topology optimization approaches that use fixed background meshes to discretize the geometry and/or the state variable fields. 
These approaches typically split elements in which the interface is embedded (\cite{van2013level,sigmund2013topology}).
The intersected elements are decomposed into sub-elements such that a face in 3D, or an edge in 2D, aligns with the interface defined by the isocontour of the $\text{LSF}$.
A solid phase needs to be assigned to each of the generated sub-elements. 
However, certain geometrical configurations suffer from ambiguous definitions of the interface, and thus the phase cannot be uniquely determined for all sub-elements in these cases.
This scenario is displayed in the zoomed-in region of the 2D embedded topology in Fig.~\ref{fig:elemAmbIssue2D}.
The level set values at the corner nodes of the center background element allow for two different interpretations of the geometry. 
Although both configurations are consistent with the discretized $\text{LSF}$, one of them creates connections between the void inclusions. 

%Figure 

In a 2D domain discretized by quadrilateral elements, ambiguous topological features exist only if an element is intersected by an isocontour more than once.
In 3D domains composed of hexahedral background elements, this issue is also found in simpler intersection configurations that describe a single interface, and is aggravated if multiple interfaces are discretized within an element (\cite{Natarajan1994,albin2016computational,Namane2018}).
Furthermore, special care may be needed for neighboring background elements that contain sub-elements with ambiguous phases assignment.
Inadequate handling of sub-elements that share common faces across adjacent background elements may introduce inconsistencies in the geometry description (\cite{Nguyen2018}).

Although ambiguous geometrical configurations are avoided if a $\text{LSF}$ is discretized using tetrahedral elements, hexahedral elements remain the preferred choice (\cite{Zhang2012}).
This is in part because hexahedral elements provide a good compromise between low computational costs and accurate approximations.
Furthermore, the need of mesh refinement is reduced in hexahedral elements with complex intersection configurations (i.e., multiple interfaces) if higher order shape functions and/or B-splines are used to adequately predict physical responses (\cite{Noel2019HB}). 

Geometrical aspects of the aforementioned ambiguities have been studied in applications ranging from medical data visualization to computer graphics problems. 
Techniques developed to resolve ambiguous topological features have been formulated for a wide range of requirements such as preserving topology, filtering noise, pleasing visual appearance, among others (\cite{Levoy1988,Yamasaki2015,laurain2018analyzing}).
Some of these techniques seek to improve continuity under continuous deformations, which can restrict the evolution of the geometry in an optimization process. 
For example, a class of methods that aim at preserving the topology perform nodal analyses to identify critical nodes that generate ambiguous configurations. 
In these techniques, only shape changes are allowed by preventing sign changes in the level set value of these critical nodes; see \cite{kong1989digital,Han2003}.
Applying them in a level set topology optimization problem would prevent the merge/split of geometric features.
Hence, not all techniques for resolving ambiguous topologies are compatible with the requirements of topology optimization.

Freedom in the evolution of a smooth topology can be achieved using methods with various degrees of complexity. 
If the generation of the interface is viewed as a boundary-value problem, the solution of a system of partial differential equations can be used to define a smooth interface (\cite{bloor1989generating, bloor1995efficient, dekanski1997partial, Chen2017}). Here, a set of parameters control the geometry through boundary conditions along the interface.
Implementing these methods introduces significant additional complexities associated with building a new system of equations and mapping the resulting surface onto the discretized domain.
A simpler approach consists of locally refining intersected subdomains such that a background element is only intersected once (\cite{Cline1988}). 
These methods reduce the complexity of the intersection configurations and improve geometric resolution.
Yet, ambiguities may still exist, and the computational framework employed requires handling hanging nodes in the case of hierarchical mesh refinement. 
As an alternative, intersected elements can be discretized using boundary conformal polyhedral instead of tetrahedra; see \cite{Nguyen2018}. 
Inter-element ambiguities in this case are resolved by the so-called asymptotic decider. This technique uses the nodal level set values of the hexahedron shared face to linearly interpolate the level set value at the saddle point; see \cite{nielson1991asymptotic} for an in-depth explanation.
Though, implementing this approach requires the use of non-traditional finite elements, i.e. polyhedral elements.
Another way to circumvent ambiguous topological features consists of subdividing intersected hexahedra into structured tetrahedra, and discretizing the $\text{LSF}$ on these generated sub-elements (e.g., marching tetrahedra \cite{doi1991efficient,zhou1995elaborate}). 
However, not only a considerably larger number of tetrahedra are required in these methods, but the approximated $\text{LSF}$ may compromise smoothness in the resulting topologies (\cite{treece1999regularised}).

Abstaining form mesh refinement methods, techniques used to resolve topological ambiguities either resort to geometric indicators or user-defined criteria for assigning phases to sub-elements.
In this case, a phase is assigned to tetrahedralized hexahedra based on measurements of area, curvature, etc., as shown in, for example, \cite{dyn2001optimizing, gatzke2006estimating, albin2016computational}; or by consistently selecting a phase for all ambiguous sub-elements.
This class of methods is explored in this paper.
 
The study presented here focuses on analyzing topological ambiguities in the description of embedded geometries from discretized $\text{LSFs}$ on hexahedral meshes.
Ambiguous topological features are categorized, and
a set of rules to determine a phase in ambiguous tetrahedral sub-elements based on either user-defined or geometric indicators is introduced.
Local (element-wise) and global (inter-element) rules are tested using different optimization formulations to assess their influence on the evolution of the geometry of the optimization process and performance of the optimized design.
Two-phase, solid-void design optimization problems are considered and are solved with an explicit level set topology optimization framework.
The extended finite element method ($\text{XFEM}$) is used to discretize the weak form of the governing equations; see \cite{belytschko2009review}.
A linear elastic behavior is considered in all structural design problems.

The remainder of the paper is organized as follows: Section \ref{sec:topAmbIn3D} details the characteristics of the geometrical configurations affected by the topological ambiguities. In Section \ref{sec:rulesForTopAmb}, a set of rules for resolving such ambiguities, as well as simple algorithms as guidelines for implementation, are provided. The level set $\text{XFEM}$ optimization framework used in this study is summarized in Section \ref{sec:LsXFEMTopOptFramework}. The influence of the rules introduced in Section \ref{sec:rulesForTopAmb} is examined with structural design problems in Section \ref{sec:NumEx}. Finally, Section \ref{sec:concl} summarizes this paper and provides directions for future work.

%-------------------------------------------------------------------------------------------------%
\section{Topology Ambiguities in Level Set Methods} \label{sec:topAmbIn3D}
%-------------------------------------------------------------------------------------------------%

\subsection{Geometry description}
%----------------------------------------%

The layout of a two-phase solid-void problem in a design domain, $\Omega$, is described by a $\text{LSF}$, $\phi$, as follows:
% Equation
\begin{equation}
\begin{aligned}
\phi(\boldsymbol X)
\begin{cases}
> 0, ~~~~~& \forall~\boldsymbol X \in \Omega_{S}, \\
< 0, ~~~~~& \forall~\boldsymbol X \in \Omega_{V} , \\
= 0, ~~~~~& \forall~\boldsymbol X \in \Gamma_{S,V}.
\end{cases}
\end{aligned}
\label{eq:LSFieldDef}
\end{equation}
The design domain is subdivided into the solid and void phases, $\Omega_{S}$ and $\Omega_{V}$, respectively, such that $\Omega_{S} \cup \Omega_{V} = \Omega$. The interface, denoted by $\Gamma_{S,V}$, is identified by the zero level set isocontour.

The $\text{LSF}$ is discretized by tri-linear shape function on a structured mesh, $\mathcal{H}$.
This mesh is composed of regular eight-node hexahedra that are classified into two groups, $\mathcal{H}_{U}$ and $\mathcal{H}_{I}$, with $\mathcal{H}_{U}\cup\mathcal{H}_{I}=\mathcal{H}$.
The subdomain $\mathcal{H}_{U}$ contains all the un-intersected hexahedra.
A hexahedron, $\mathcal{H}_i$, is part of this set if all their nodal level set values are either positive or negative.
Furthermore, the subset $\mathcal{H}_{I}$ consists of intersected hexahedra defined as all hexahedra for which at least one of their eight nodal level set values differs in sign from the remaining nodal level set values.

All intersected hexahedral elements, i.e. $\mathcal{H}_i \in \mathcal{H}_{I}$, are split into tetrahedral sub-elements via a tetrahedralization algorithm. 
The vertices of the tetrahedra are constructed from vertices of the background element and intersection points along the edges. 
Hence, some of the faces of these generated tetrahedra contain intersection points only.
Tetrahedralization algorithms employed for this subdivision receive the nodal level set values and the intersection points of a hexahedron $\mathcal{H}_i$ as input and return the geometrical information of the set of generated tetrahedra, $\mathcal{T}^{\mathcal{H}_i}$.
In this work, all the generated tetrahedra are contained in the set $\mathcal{T}= \{\mathcal{T}^{\mathcal{H}_i}:\mathcal{H}_i \in \mathcal{H}_{I}\}$. 
The Delaunay method (\cite{lee1980two,shewchuk2002delaunay}) is used for tetrahedralizing the intersected hexahedral elements in this study.

\begin{figure}[h]
	\centering
	\includegraphics[width=1.0\linewidth]{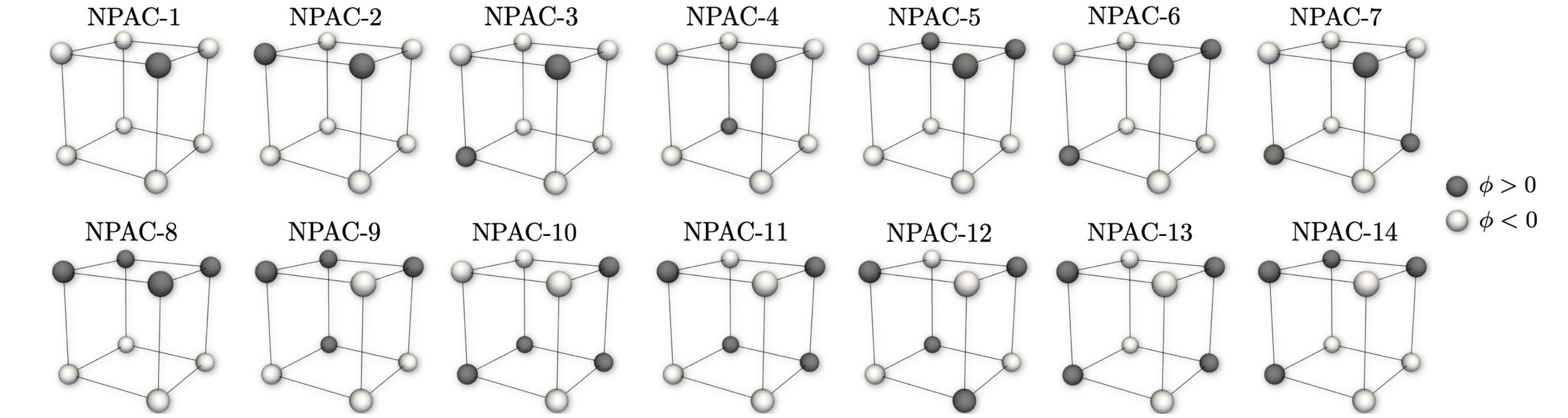}
	\caption{All unique $\text{NPACs}$ using a discretized $\text{LSF}$ onto an intersected eight-node hexahedron.}
	\label{fig:nodalLSValueAssignmentAllCases}
\end{figure}

The intersection points along the edges are defined by the points with a zero level set value. 
The level set values along an edge are linearly interpolated from the level set values of the edge's corner nodes.
Thus, the linear interpolation of the $\text{LSF}$ limits the possible configurations to one intersection per edge.
Since each node can have a positive or negative level set value, a total of $2^8=256$ nodal phase assignment configurations ($\text{NPACs}$) are possible in an eight-node hexahedral element. 
After accounting for reflective and rotational symmetries, the number of unique $\text{NPACs}$ for intersected elements is reduced to the 14 cases shown in Fig.~\ref{fig:nodalLSValueAssignmentAllCases}.
The reader is referred to \cite{Natarajan1994,albin2016computational,Namane2018} for a case-by-case analysis, as well as detailed visual schematics.

%Figure 

\begin{figure}[h!]
	\centering
	\includegraphics[width=1.0\linewidth]{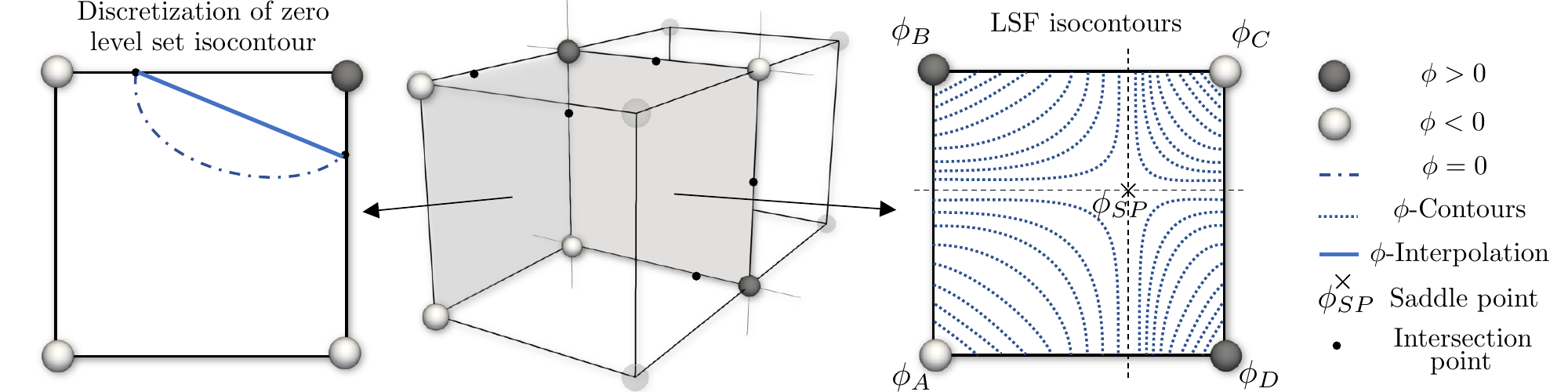}
	\caption{Two contiguous hexahedra which discretized geometry present ambiguous phase assignment (center). The highlighted faces show (left) the difference between approximated level set field and discretized geometry, and (right) level set field contours on such face and its saddle point.}
	\label{fig:asymptoticDeciderExplanation}
\end{figure}

In this work, the discretized geometry is constructed by connecting intersection points along intersected edges. 
Thus, a small discrepancy between the discretized geometry and the approximated level set field using linear interpolation functions is expected. 
The left side of Fig.~\ref{fig:asymptoticDeciderExplanation} illustrates this in a simple 2D schematic. 
It can be seen that the zero isocontour of the $\text{LSF}$, described by a blue dashed curve, is approximated by a straight blue line between two intersected edges.

%Figure 

Once intersected hexahedra have been split into tetrahedra, one or multiple interfaces must be defined to approximate the problem's geometry.
This implies that either a solid or a void phase must be assigned to every tetrahedron, $\mathcal{T}_i \in \mathcal{T}$.
In some cases, however, one or more tetrahedra can be defined as solid or void while keeping consistency with the associated $\text{NPAC}$.
The lack of uniqueness inherent in such cases leads to multiple possible interface definitions.
In other words, different discretized domains can be generated from the same discretized $\text{LSF}$ depending on the phase assigned to these ambiguous tetrahedra, as shown in Fig.~\ref{fig:elemAmbIssue2D} in 2D.

\subsection{Ambiguous tetrahedra ($\text{ATs}$)} \label{subsec:AmbTet}
%----------------------------------------------------%
An ambiguous tetrahedron, $\text{AT}$, is defined as any tetrahedron whose vertices are all intersection points.
Hence, a minimum of four intersection points is needed for an $\text{AT}$ to exist. For this reason, all but the $\text{NPAC-1}$ configuration can contain ambiguities; see Fig.~\ref{fig:nodalLSValueAssignmentAllCases}.
Despite the number and shape of the generated tetrahedra may change depending on the tetrahedralization algorithm employed (\cite{shewchuk2002delaunay,xue2004reconstruction}), $\text{ATs}$ cannot be avoided.

%Figure 
\begin{figure}[h!]
	\centering
	\includegraphics[width=0.9\linewidth]{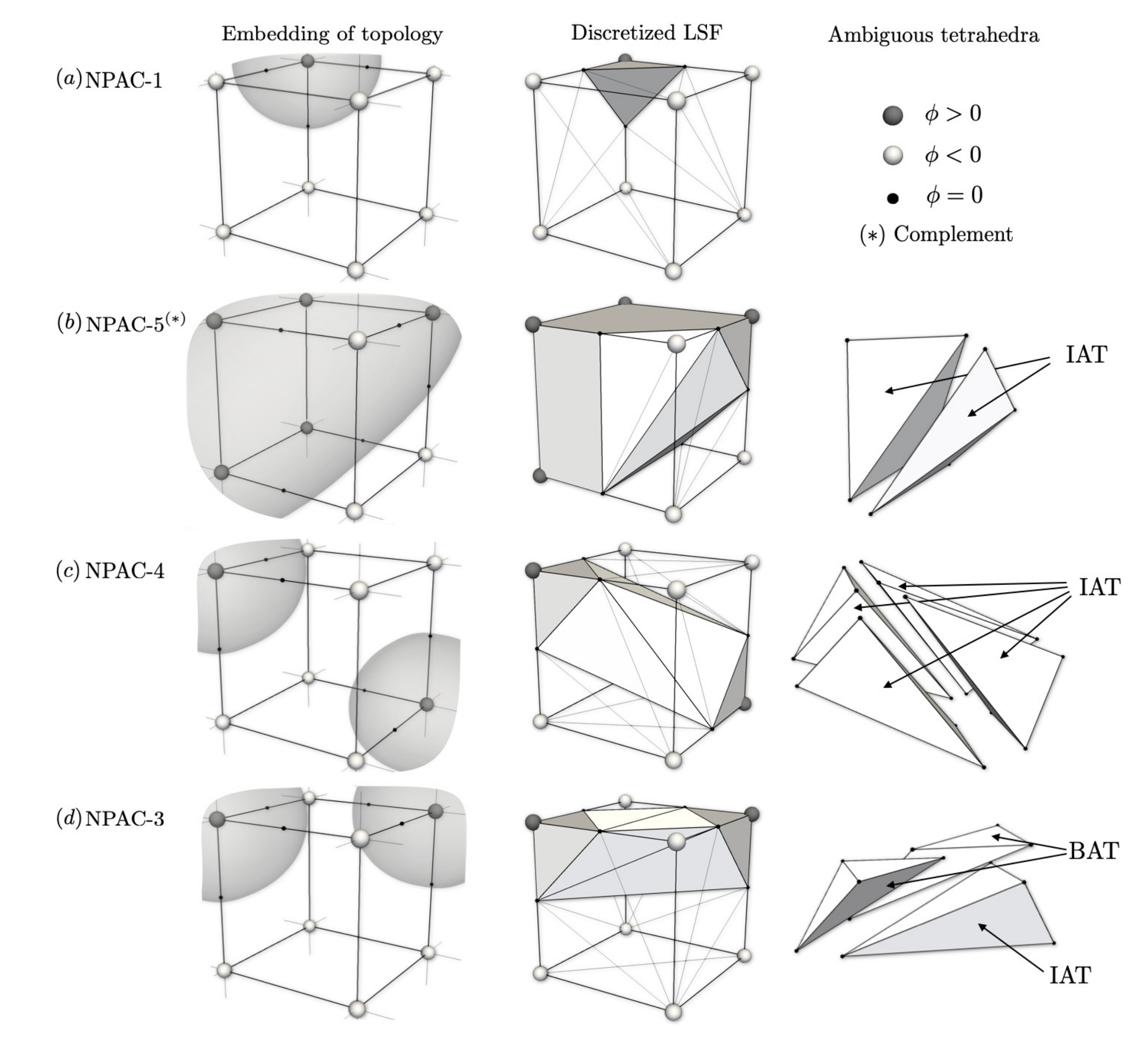}
	\caption{Illustrations of the four cases of intersection configurations for a tetrahedralized hexahedron: (a) no ambiguous sub-elements; (b) $\text{IATs}$ with one interface; (c) $\text{IATs}$ with multiple interfaces; (d) $\text{IATs}$ and $\text{BATs}$ with multiple interfaces.}
	\label{fig:hexesWithAmbSubElemsBothCases}
\end{figure}

To simplify the analysis of topology ambiguities, the $\text{NPACs}$ described in Fig.~\ref{fig:nodalLSValueAssignmentAllCases} are categorized as shown in Fig.~\ref{fig:hexesWithAmbSubElemsBothCases}.
Examples of continuous $\text{LSF}$s to be embedded in a discretized domain are shown on the left column. 
The discretized geometries are shown in dark gray, together with the $\text{ATs}$ highlighted in light gray in the center column. 
The right column shows only the $\text{ATs}$ for better visualization.
In the first case, illustrated in Fig.~\ref{fig:hexesWithAmbSubElemsBothCases}(a), no $\text{ATs}$ are present since only three intersection nodes exist.
Fig.~\ref{fig:hexesWithAmbSubElemsBothCases}(b) shows the most innocuous $\text{NPAC}$ with $\text{ATs}$, which occurs in hexahedra with a single interface.
In this configuration, assigning different phases to the $\text{ATs}$ results in different shapes but not topologies. 
If the background element is intersected more than once, as shown in Fig.~\ref{fig:hexesWithAmbSubElemsBothCases}(c), then $\text{ATs}$ can connect/disconnect interfaces, and thus alter the topology.
Note that in both Figs. \ref{fig:hexesWithAmbSubElemsBothCases}(b) and (c), none of $\text{ATs}$ faces is part of the background hexahedron faces. 
We classify them as Internal Ambiguous Tetrahedra, herein labelled as $\text{IATs}$.
However, $\text{NPACs}$ can also include $\text{ATs}$ with one of their faces overlapped with one of the hexahedron's faces, as is depicted in Fig.~\ref{fig:hexesWithAmbSubElemsBothCases}(d).
We term these Boundary Ambiguous Tetrahedra as $\text{BATs}$.

\subsubsection{Internal ambiguous tetrahedra ($\text{IATs}$)}
%---------------------------------------------------%
A key characteristic of this type of ambiguities is that at least one of the faces of an $\text{IAT}$ lays on an interface. 
If the $\text{NPAC}$ leads to only one interface, all ambiguous sub-elements are IATs, as shown in Fig.~\ref{fig:hexesWithAmbSubElemsBothCases}(b). 
Furthermore, the difference between the interface geometry approximated by the IATs and the one described by the linearly interpolated $\text{LSF}$ is typically negligible.

\subsubsection{Boundary ambiguous tetrahedra  ($\text{BATs}$)}
%----------------------------------------%
This type of $\text{AT}$ can only exist in intersected hexahedra that are able to describe multiple interfaces. Furthermore, $\text{BATs}$ are always accompanied by $\text{IATs}$.
A well-known issue of $\text{NPACs}$ with $\text{BATs}$ is guaranteeing consistency in phase assignment across two adjacent hexahedral background elements that share a face on which a $\text{BAT}$ is defined; see \cite{nielson1991asymptotic,Nguyen2018}. The sketch at the center of Fig.~\ref{fig:asymptoticDeciderExplanation} shows this scenario.
To achieve a consistent phase assignment, $\text{BATs}$ sharing a hexahedron face must have the same phase.

\subsubsection{Effect of ambiguous tetrahedra}
%----------------------------------------%

Internal ambiguities are ubiquitous and, in most cases, harmless since they do not alter significantly the geometry.
However, hexahedral background elements with multiple interfaces can have $\text{IATs}$ which phase assignment may have a strong influence on the topology of a given design. In most of those scenarios, the analysis has to include $\text{BATs}$.

\begin{figure}[h!]
	\centering
	\includegraphics[width=1.0\linewidth]{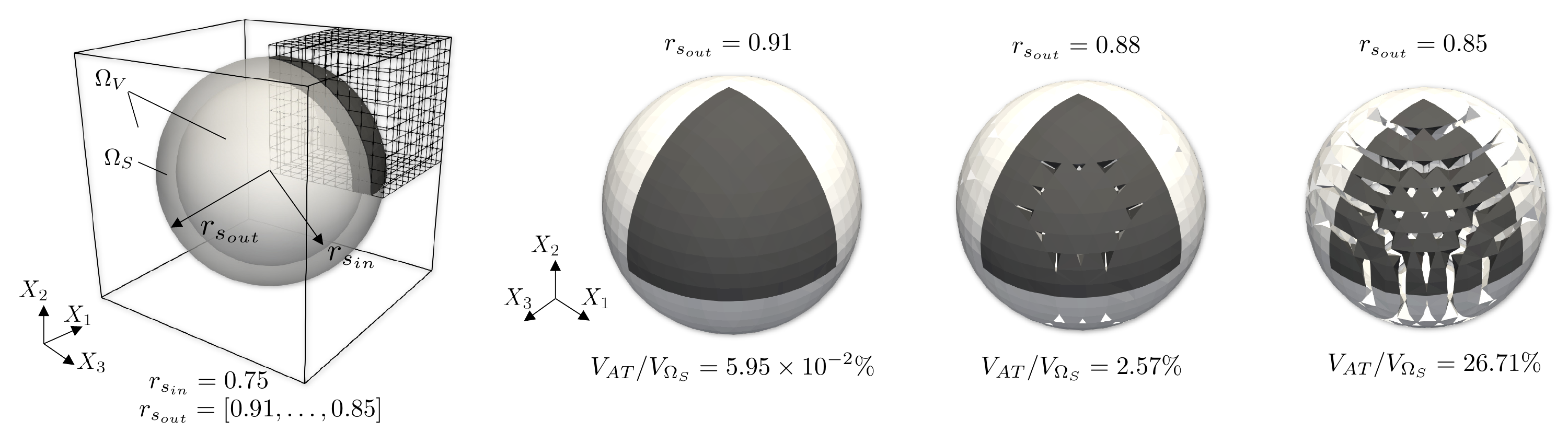}
	\caption{Spherical shell with varying thickness in an immersed boundary geometrical description. An eighth of the domain is highlighted in dark gray to enhance contrast.}
	\label{fig:sphericalShellIntAndExtAmb}
\end{figure}

Inappropriate treatment of ambiguous configurations that combine $\text{IATs}$ and $\text{BATs}$ can result in undesired geometries, especially when thin features are present in the design. Shell-like structures, which are not uncommon in engineering design problems, can be greatly affected since the appearance of dents may be favored. 
This is illustrated in the example in Fig.~\ref{fig:sphericalShellIntAndExtAmb}. 
Here, the influence of $\text{ATs}$ on the geometry representation of a spherical shell is shown for three different wall thicknesses. The shell has a constant inner radius and its thickness is reduced by varying the external radius.
An eighth of the $3\times 3\times 3$ domain is discretized by a $10\times 10\times 10$ hexahedral background mesh and is highlighted in dark gray for visualization purposes.
In this example, void phase is assigned to all $\text{ATs}$.
The discretized geometry, together with the ratio between volumes of $\text{AT}$ and solid subdomain, are provided for the three configurations on the right side of Fig.~\ref{fig:sphericalShellIntAndExtAmb}.
It can be observed that once the difference between the inner and outer radii is smaller than the edge length of a background element, the volume of the ambiguous subdomain increases up to 26.71\% of the solid subdomain.
Note that the extreme misrepresentation of the spherical shell geometry can be avoided by assigning a solid phase to all $\text{ATs}$. 
However, assigning the same phase to $\text{ATs}$ is not necessarily the most beneficial option for all intersection configurations, as is shown in Section \ref{sec:NumEx}. 

To assess whether it is more beneficial in the optimization process to use a simple criterion, such as the one used for the spherical shell, or geometrical information, a group of rules that systematically assigns a phase to $\text{AT}$ is presented in next section.

%Figure

%-------------------------------------------------------------------------------------------------%
\section{Rules for Resolving Topology Ambiguities}\label{sec:rulesForTopAmb}
%-------------------------------------------------------------------------------------------------%

The ambiguities described in Section \ref{subsec:AmbTet} can be resolved by applying phase assignment rules either locally (i.e., separately for each hexahedral background element) or globally (i.e., on groups of adjacent hexahedra). 
The local and global rules explored in this paper are denoted by $\mathcal{L}$ and $\mathcal{G}$, respectively.
These rules collect a subset of techniques that allow for the free evolution of the geometry in a shape or topology optimization process, but this subset is by no means complete. The reader is referred to \cite{Zhang2012,nielson1991asymptotic,zhou1995elaborate,gatzke2006estimating} for more information on rules used to this end.

%-----------------------------------------------------------%
\subsection{Local rules ($\mathcal{L}$)}
%-----------------------------------------------------------%

The local rules described in this section resolve $\text{IATs}$ and $\text{BATs}$ separately, using one criterion for each ambiguity type.
Consistent phase assignment across hexahedra (i.e., $\text{BAT}$ phase assignment) is achieved by enforcing the asymptotic decider \cite{nielson1991asymptotic}. 
This technique uses the nodal level set values of the hexahedron face containing a face of a $\text{BAT}$ to compute the level set value at the saddle point, $\phi_{sp}$, assuming a bi-liner interpolation of the level set field over the hexahedron face. 
The right side of Fig.~\ref{fig:asymptoticDeciderExplanation} shows a schematic of the location of $\phi_{sp}$ on a face that contains $\text{BATs}$.
The level set value at the saddle point is computed as follows:
% Equation
\begin{equation}
\begin{aligned}
\phi_{sp} = \frac{\phi_{A}\phi_{C} - \phi_{B}\phi_{D}}{\phi_{A}+\phi_{B}+\phi_{C}+\phi_{D}},
\end{aligned}
\label{eq:asymDecEq}
\end{equation}
where $\phi_{A}$, $\phi_{B}$, $\phi_{C}$, and $\phi_{D}$ correspond to the nodal level set values of the sharing hexahedron face.
The sign of $\phi_{sp}$ determines the phase to be assigned. 
If $\phi_{sp} \geq 0$, then all $\text{BATs}$ in the adjacent hexahedra sharing the face are assigned the solid phase. Conversely, if $\phi_{sp} < 0$, $\text{BATs}$ are assigned the void phase.
The reader is referred to \cite{nielson1991asymptotic} for details of this technique. 

After $\text{BATs}$ have been resolved via the asymptotic decider, the phase of the remaining $\text{IATs}$ is determined using one of the rules described below (in order of increasing complexity in terms of implementation).

\subsubsection{Fixed phase assignment ($\mathcal{L}_1^{\Omega_{S}}$/$\mathcal{L}_1^{\Omega_{V}}$)}
% ---------------------------------------------------------------
In this rule, all $\text{IATs}$ are assigned the same phase throughout the optimization process. 
If ambiguous sub-elements are assigned the solid phase, the rule is denoted as $\mathcal{L}_1^{\Omega_{S}}$; conversely, if void phase is assigned, it is named $\mathcal{L}_1^{\Omega_{V}}$.
This approach requires no additional computations for phase determination, as can be seen in Algorithm \ref{alg:L1Scheme}.

% Algorithm
\begin{algorithm}[h] \label{alg:L1Scheme}
\SetNoFillComment
\caption{Resolving $\text{IATs}$ using the $\mathcal{L}_1$ scheme}
{
// Loop over all intersected hexahedra of the background mesh \\ 
\For{$\mathcal{H}_i \in \mathcal{H}_\mathcal{I}$}
{
// Check if any of the generated tetrahedral sub-elements are ambiguous \\ 
\uIf {${IAT} \neq \{\}$}
{ 
// Assign predefined phase to all $\text{IAT}$ \\
$p=\Omega_{S}/\Omega_{V}$ for $\mathcal{L}_1^{\Omega_{S}}/\mathcal{L}_1^{\Omega_{V}}$
}
}
}
\end{algorithm}

\subsubsection{Previous main phase in optimization process ($\mathcal{L}_2$)}
% -----------------------------------------------------------------------------------------
An easy-to-implement alternative that promotes a smoother evolution of the design during the optimization process is choosing the phase assigned to the $\text{IATs}$ in the current design iteration, $\mathcal{D}_{it}$, based on an elemental phase, $p_{\mathcal{H}_i}$, function of the previous design iteration, $\mathcal{D}_{it-1}$. 
If, for example, a hexahedron with the configuration $\text{NPAC-4}$ shown in Fig.~\ref{fig:hexesWithAmbSubElemsBothCases}(b) is part of the void phase at $\mathcal{D}_{it-1}$, this rule prevents connecting the two interfaces by setting the $\text{IATs}$ to void phase.
In this scheme, all internal ambiguities are assumed to be solid phase in the first design iteration (i.e., at $\mathcal{D}_{it}=1$) to promote a topology preserving evolution of the design at the initial stage of the optimization process.
An implementation of this logic is detailed in Algorithm \ref{alg:L2Scheme}. 

% Algorithm
\begin{algorithm}[h] \label{alg:L2Scheme}
\SetNoFillComment
\caption{Resolving $\text{IATs}$ using the $\mathcal{L}_2$ scheme}
// Loop over all hexahedra of the background mesh \\ 
\For{$\mathcal{H}_i \in \mathcal{H}$}
{
// Check if element is not intersected to initialize elemental phase $p_{\mathcal{H}_i}$\\ 
\eIf {$\mathcal{H}_i \in \mathcal{H}_\mathcal{U}$} 
{ 
// Store phase of un-intersected hexahedron \\ 
$p_{\mathcal{H}_i}=\Omega_{S} \vee p_{\mathcal{H}_i}=\Omega_{V}$
}
({$(\mathcal{H}_i \in \mathcal{H}_\mathcal{I})$})
{
\uIf {$\mathcal{D}_{it} = 1$} 
{ 
// If this is the first design iteration, set elemental phase to solid phase  \\
$p_{\mathcal{H}_i}=\Omega_{S}$
}
// Assign phase $p_{\mathcal{H}_i}$ to the ambiguous sub-elements\\
    \uIf{${IAT} \neq \{\}$}
    {
    $\text{IATs}$ phase  = $p_{\mathcal{H}_i}$ \\
    }
}
}
\end{algorithm}

\subsubsection{Based on level set value at averaged centroid ($\mathcal{L}_3$)}
\begin{figure}[h!]
	\centering
	\includegraphics[width=1.0\linewidth]{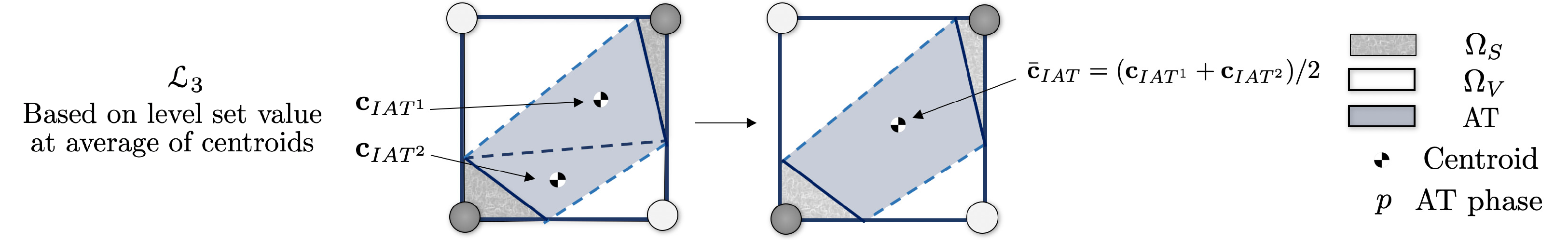}
	\caption{2D schematics of the $\mathcal{L}_3$ rule.}
	\label{fig:AvgCentroidRule_L3}
\end{figure}

% ---------------------------------------------------------------
Expanding the idea of the asymptotic decider to 3D, $\text{IATs}$ can be resolved based on the sign of a single interpolated level set value in $\mathcal{H}_{i}$. 
The decider point (equivalent to the saddle point of the asymptotic decider) is defined as the average of the centroids of the $\text{IAT}$ within a background element.
This is illustrated in 2D in Fig.~\ref{fig:AvgCentroidRule_L3}, where the centroid of the ambiguous subdomain, $\bar{\mathbf{c}}_{IAT}$, is used to determine a phase.
Note that information of the previous optimization step is not needed in this rule.
Also, in hexahedra with a single interface, the phase is assigned in accordance to the unambiguous $\text{LSF}$ used to construct the interface. 
A scheme that implements this rule is shown in Algorithm \ref{alg:L3Scheme}.

%Figure 

% Algorithm
\begin{algorithm}[h!] \label{alg:L3Scheme}
\SetNoFillComment
\caption{Resolving $\text{IATs}$ using the $\mathcal{L}_3$ scheme}
// Loop over all intersected hexahedra of the background mesh \\ 
\For{$\mathcal{H}_i \in\mathcal{H}_{\mathcal{I}}$}
{
    // Check if hexahedron contains ambiguous sub-elements \\
    \uIf{${IAT} \neq \{\}$}
    {
    // Compute centroids of every $\text{IAT}$ using the coordinates of their vertices, $\mathbf{X}_{IAT^{v_i}}$\\
    $\mathbf{c}_{IAT} = \displaystyle\sum_{i=1}^{N_v=4} \mathbf{X}_{IAT^{v_i}} / N_v $ \\
    // Compute the average of the centroids of the $N_{IAT}$ ambiguous sub-elements \\
    $\bar{\mathbf{c}}_{IAT} = \displaystyle\sum_{i=1}^{N_{IAT}} \mathbf{c}_{IAT_i}  / N_{IAT}$ \\
    // Interpolate the level set value at centroid using linear shape functions, $\boldsymbol{\mathcal{N}}$, and the nodal level set values, $\hat{\boldsymbol{\phi}}^{\mathcal{H}_i}$, of the containing hexahedron \\
    $\phi_{\bar{c}_{IAT}} = \boldsymbol{\mathcal{N}}(\bar{\mathbf{c}}_{IAT}) \hat{\boldsymbol{\phi}}^{\mathcal{H}_i} $  \\
    // Assign elemental phase based on the sign of the level set value \\
    \textbf{if} ( $\phi_{\bar{c}_{{IAT}}} \geq 0 )~~ p_{\mathcal{H}_i} = \Omega_{S} ~~ \textbf{else}~~ p_{\mathcal{H}_i} = \Omega_{V}$ \\
// Assign a phase to the ambiguous sub-elements\\
 $\text{IATs}$ phase  = $p_{\mathcal{H}_i}$ \\
    }      
}
\end{algorithm}

\subsubsection{Shared area between $\text{IAT}$ and neighboring tetrahedra per phase ($\mathcal{L}_4^{A_{+}}$/$\mathcal{L}_4^{A_{-}}$)} \label{subsec:L4Rule}
\begin{figure}[h!]
	\centering
	\includegraphics[width=1.0\linewidth]{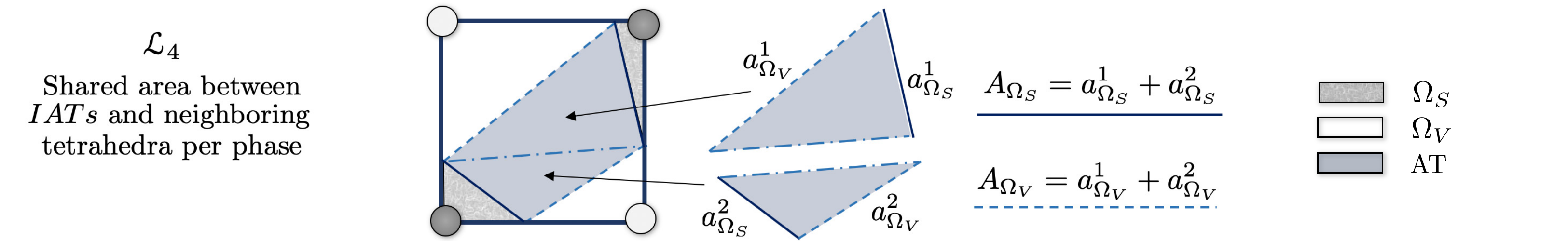}
	\caption{2D schematics of the $\mathcal{L}_4$ rule.}
	\label{fig:AvgCentroidRule_L4}
\end{figure}

% ---------------------------------------------------------------
To avoid interpolating the $\text{LSF}$ to resolve the $\text{IATs}$, a rule based on a geometric indicator is proposed.
In this rule, the shared surface area between each $\text{IAT}$ and its unambiguous neighboring tetrahedra are computed per phase.
The shared surface areas of two neighboring $\text{IATs}$ are omitted from this calculation.
The phase of the unambiguous tetrahedra with whom the $\text{IATs}$ share the maximum ($\mathcal{L}_4^{A_{+}}$) or minimum ($\mathcal{L}_4^{A_{-}}$) surface is chosen for all $\text{IAT}$. 
Figure~\ref{fig:AvgCentroidRule_L4} shows a 2D schematic of this rule.

Note that implementing this scheme requires knowledge of the phase of adjacent tetrahedra. 
Hence, the connectivity of the generated tetrahedra must be constructed per intersected hexahedron.
An approach that employs this rule is summarized in Algorithm \ref{alg:L4Scheme}.

%Figure 

% Algorithm
\begin{algorithm}[h!] \label{alg:L4Scheme}
\SetNoFillComment
\caption{Resolving $\text{IATs}$ using the $\mathcal{L}_4$ scheme}
// Loop over all intersected hexahedra of the background mesh \\ 
\For{$\mathcal{H}_i \in\mathcal{H}_{\mathcal{I}}$}
{
// Check if hexahedron contains ambiguous sub-elements \\
\uIf{${IAT} \neq \{\}$}
{
       // Initialize areas for both phases \\
       $A_{\Omega_{S}}= A_{\Omega_{V}}= 0$ \\ 
       // Loop over all ambiguous tetrahedra in intersected hexahedron\\
       \For{ $\mathcal{T}_j^{\mathcal{H}_i} \in {IAT}$ }
       {
           // Loop over the faces, $\mathcal{F}_k$, of current ambiguous tetrahedron \\
            \For{$\mathcal{F}_k \in \mathcal{T}_j^{\mathcal{H}_i}$}
	    {
	    // Get tetrahedra sharing current face \\
	    $\mathcal{T}_{\mathcal{F}^1_k }^{\mathcal{H}_i}$ and $\mathcal{T}_{\mathcal{F}^2_k }^{\mathcal{H}_i}$ sharing $\mathcal{F}_k$\\ 
	    // Skip face computations if both tetrahedra are ambiguous. \\
	\eIf{ ($\mathcal{T}_{\mathcal{F}^1_k }^{\mathcal{H}_i} ~\&~ \mathcal{T}_{\mathcal{F}^2_k }^{\mathcal{H}_i}$) $ \in IATs$ }
	{
	    continue
	}
	{
	    // Compute face area, $A_{\mathcal{F}_i}$ \\
	    // Update area of shared tetrahedra based on the phase of unambiguous tetrahedron\\
	    $(A_{\Omega_S} = A_{\Omega_S} + A_{\mathcal{F}_i})$ $\vee$ $(A_{\Omega_V} = A_{\Omega_V} + A_{\mathcal{F}_i})$ \\
	    }
	    }
	}
	// Assign elemental phase based on shared area \\
	\eIf{ $A_{\Omega_{S}} > A_{\Omega_{V}}$ }
	{
	    $p_{\mathcal{H}_i} =\Omega_{S}$ for $\mathcal{L}_4^{A_{+}}$, or $p_{\mathcal{H}_i} =\Omega_{V}$ for $\mathcal{L}_4^{A_{-}}$ 
	}
	{
	    $p_{\mathcal{H}_i} =\Omega_{V}$ for $\mathcal{L}_4^{A_{+}}$, or $p_{\mathcal{H}_i} =\Omega_{S}$ for $\mathcal{L}_4^{A_{-}}$ 
	} 
// Assign a phase to the ambiguous sub-elements\\
$\text{IATs}$ phase  = $p_{\mathcal{H}_i}$ \\
}
}
\end{algorithm}

%-----------------------------------------------------------%
\subsection{Global rules for resolving topology ambiguities ($\mathcal{G}$)}
%-----------------------------------------------------------%

In this set of rules, the asymptotic decider is no longer used to enforce consistent phase assignment to $\text{BATs}$ across intersected hexahedra. 
Here both $\text{IAT}$ and $\text{BAT}$ are resolved using the same criterion. Consequently, chains of connected $\text{ATs}$ across multiple elements can be formed.
The criteria described below are applied either on the entire domain ($\mathcal{G}_1$) or per cluster of connected ambiguous tetrahedra connected across intersected tetrahedra ($\mathcal{G}_2$).

\subsubsection{Fixed phase assignment ($\mathcal{G}_1^{\Omega_{S}}$/$\mathcal{G}_1^{\Omega_{V}}$)} \label{subsec:gblFxdPhaseAssigned}
%-----------------------------------------------------------%
This rule enforces either solid ($\mathcal{G}_1^{\Omega_{S}}$), or void ($\mathcal{G}_1^{\Omega_{V}}$) phase to all ambiguous tetrahedra.
Similar to $\mathcal{L}_1$, no additional implementation is needed once the $\text{ATs}$ have been identified. However, in contrast to $\mathcal{L}_1$, here $\text{BATs}$ are also included.
Note that for this rule the choice of using $\mathcal{G}_1^{\Omega_{S}}$ or  $\mathcal{G}_1^{\Omega_{V}}$ may have a major impact on the resulting geometry compared to the other rules presented in this paper.

\subsubsection{Shared area between ambiguous and uniquely defined tetrahedra per phase ($\mathcal{G}_2^{A_{+}}$/$\mathcal{G}_2^{A_{-}}$)} 
%-----------------------------------------------------------%
This global rule is an extension of the $\mathcal{L}_4$ local rule and is enforced on clusters of connected tetrahedra. These clusters can exist across intersected hexahedra instead of being contained within a single hexahedron; see Section \ref{subsec:L4Rule}.
Hence, knowledge of the connectivity of all tetrahedra, which changes for every design iteration, is needed to identify these clusters.
A graph search scheme to generate this connectivity that avoids issues associated with memory and/or computational time consumption is provided in Algorithm \ref{alg:G2Scheme}.
Note that implementing this rule is laborious, especially in parallelized software. 
Existing third party libraries, such as MOAB (\cite{tautges2004moab}) or the STK package that is part of Trilinos (\cite{Trilinos-Overview}), provide robust frameworks to manage mesh databases and can be used to ease the search of connected $\text{ATs}$. 

% Algorithm
\begin{algorithm}[h!] \label{alg:G2Scheme}
\SetNoFillComment
\caption{Resolving $\text{ATs}$ using the $\mathcal{G}_2$ scheme}
// Initialize container of clusters,
$\boldsymbol{\mathcal{C}}$ = [] \\    
// Initialize flag for $N_{\mathcal{T} \in \mathcal{H}_{\mathcal{I}}}$ tetrahedra of intersected background elements, $\boldsymbol{\mathcal{P}}$:
${\mathcal{P}}_i$ = false, $i=1,...,N_{\mathcal{T} \in \mathcal{H}_{\mathcal{I}}}$\\  
// Loop over all intersected hexahedra of the background mesh \\ 
\For{All tetrahedra faces, $\mathcal{F}^{\mathcal{T} \in \mathcal{H}_{\mathcal{I}}}$}  
{
    // Get tetrahedra sharing current face, $\mathcal{F}_i$, \\
    // Check if any tetrahedra sharing $\mathcal{F}_i$ is ambiguous and has not been processed \\
    \uIf{($\mathcal{T}_{\mathcal{F}_i} \in IAT$) \& ($\mathcal{P}_{\mathcal{T}_{\mathcal{F}_i}}=\text{false}$) }
    {
    // Initialize a face list with current face, $\boldsymbol{\mathcal{C}^{\mathcal{F}}} = [\mathcal{F}_i]$  \\ 
    // Initialize an empty cluster of $\text{AT}$, $\boldsymbol{\mathcal{C}^{AT}}$  = []\\ 
    // Process tetrahedra in face list \\
    \While{$\boldsymbol{\mathcal{C}^{\mathcal{F}}} \neq []$}
    {
    	// Loop over tetrahedra sharing current face, $\mathcal{T}_j \in \mathcal{F}_i$ \\
        \For{$\mathcal{T}_j \in \mathcal{F}_i$}
        {
        // If tetrahedron is ambiguous, process it \\
        \uIf{$\mathcal{T}_j \in AT$}
	{
	// Add tetrahedron to current cluster \\ 
	$\boldsymbol{\mathcal{C}^{AT}}$  = [ $\boldsymbol{\mathcal{C}^{AT}}$, $\mathcal{T}_j$] \\
	// Loop over faces in current tetrahedron, $ \mathcal{F}_k \in \mathcal{T}_j$ \\
	\For{$ \mathcal{F}_k \in \mathcal{T}_j$ }
        {
        // Add faces to list \\
        $\boldsymbol{\mathcal{C}^{\mathcal{F}}} = [\boldsymbol{\mathcal{C}^{\mathcal{F}}}, \mathcal{F}_k]$ 
        }
        // Flag tetrahedron as processed to avoid repeating searches in first loop \\
        $\mathcal{P}_{\mathcal{T}_j}=\text{true}$
        }
        }
        // Remove current face in $\boldsymbol{\mathcal{C}^{\mathcal{F}}}$ list
    }
}
    // If a new cluster was generated from the processed tetrahedra \\
    \uIf{$\boldsymbol{\mathcal{C}^{AT}} \neq []$}
    {
    // Add new cluster to container of clusters \\
    $\boldsymbol{\mathcal{C}}$ = [$\boldsymbol{\mathcal{C}}$,  $\boldsymbol{\mathcal{C}^{AT}}$]\\    
    }
}
// After all clusters of connected $\text{ATs}$ have been generated, loop over them \\
\For{$\boldsymbol{\mathcal{C}^{AT}}_i \in \boldsymbol{\mathcal{C}}$}  
{
// Initialize areas for both phases: $A_{\Omega_{S}}= A_{\Omega_{V}}= 0$ \\ 
// Loop over faces of current cluster \\
\For{$\mathcal{F}_{j} \in \boldsymbol{\mathcal{C}^{{AT}}}_i$}  
{
    // Get tetrahedra sharing current face, $\mathcal{T}_k \in \mathcal{F}_j$ \\
    // Check if only one of the shared tetrahedra is ambiguous \\
\uIf{Only one $\mathcal{F}_j \in AT $ }
{
	    // Compute face area, $A_{\mathcal{F}_j}$ \\
	    // Update area of phase of shared tetrahedra (either $p=\Omega_{S}$ or $p=\Omega_{V}$) \\
	    $A_{p} = A_{p} + A_{\mathcal{F}_j}$ \\
}
}
	// Assign phase based on shared area \\
	\eIf{ $A_{\Omega_{S}} > A_{\Omega_{V}}$ }
	{
$p=\Omega_{S}$ for $\mathcal{G}_2^{A_{+}}$, or $p=\Omega_{V}$ for $\mathcal{G}_2^{A_{-}}$ 
	}
	{
$p=\Omega_{V}$ for $\mathcal{G}_2^{A_{+}}$, or $p=\Omega_{S}$ for $\mathcal{G}_2^{A_{-}}$ 
	} 
	// Loop over all ambiguous sub-elements in current cluster \\
\For{$\mathcal{T}_j$ in $\mathcal{C}^{{AT}}_i$}  
{
Assign phase $p$ \\
}
}
\end{algorithm}

%-----------------------------------------------------------%
\subsection{Local versus global rules}
%-----------------------------------------------------------%

In terms of implementation, local rules $\mathcal{L}_1$, $\mathcal{L}_2$ and $\mathcal{L}_3$, and global rule $\mathcal{G}_1$, are more attractive as they require little effort. 
Additionally, these rules have a low computational overhead.
However, rules $\mathcal{L}_4$ and $\mathcal{G}_2$ provide a more consistent/smooth geometry description at the cost of higher computational effort.
The advantage of these last two rules comes at the cost of involved implementations of problem specific geometric indicators to resolve ambiguities.

Local rules $\mathcal{L}_1$, $\mathcal{L}_2$ and $\mathcal{L}_3$ allow for imperfections in the geometry description due to disparity in phase assignment between $\text{BAT}$ and $\text{IAT}$.
Either solid floating pieces embedded in a mostly void subdomain, or small holes in a mostly solid subdomain, are likely to occur in hexahedra with multiple interfaces, see \cite{treece1999regularised}.
In the former case, the numerical technique used for the physical analysis must include appropriate treatment of these disconnected pieces, such as consistent generalized enrichment when using $\text{XFEM}$ and adding stiffness to avoid zero energy modes; see Section \ref{subsec:XFEMTheory}.
The local rule $\mathcal{L}_4$ does not suffer from the issues described above since the phase of the interior clusters is determined from the phase of the surrounding unambiguous tetrahedra.

Global rules have a more pronounced effect on the resulting discretized geometry compared to local rules. 
Although only rarely observed, assigning different phases to large clusters of connected $\text{ATs}$ can significantly alter the topology.
Differences in topologies are less evident if local rules are used because of the asymptotic decider. 
The influence (if any) of using the rules described in this section on the smoothness of the evolution of the geometry in a topology optimization process, as well as the performance of the optimized designs, has been unexplored to date. 
In this paper, a study of these rules is presented in Section \ref{sec:NumEx}.

%-------------------------------------------------------------------------------------------------%
\section{Explicit Level Set Topology Optimization Framework} \label{sec:LsXFEMTopOptFramework}
%-------------------------------------------------------------------------------------------------%

The design, i.e. the geometry, is controlled by a vector of design variables, 
${\boldsymbol s} := \{ {\boldsymbol s} \in \rm I\!R^{N_s} |~ \phi_{low} \leq {s}_i \leq \phi_{up}, i=1,...,N_s \}$ bounded between user-defined lower, $\phi_{low}$, and upper, $\phi_{up}$, bounds.
A design variable is assigned to each node of the structured hexahedral background mesh; thus, $N_s$ equals the number of nodes of the background mesh; see \cite{van2013level,sigmund2013topology} for details.

To increase numerical stability and enhance convergence of the optimization problem, the explicit relationship between the design variables and the nodal coefficients of the discretized $\text{LSF}$ is defined by the linear filter described in \cite{kreissl2012levelset}. This distance-based filter is defined as:
% Equation
\begin{equation}
\begin{aligned}
\phi_i 
= \frac
{
1
}
{
\displaystyle\sum_{j=1}^{N_{r_f}} w_{ij}
} \displaystyle\sum_{j=1}^{N_{r_f}} w_{ij} s_j, 
~~~~~
w_{ij} = max(0,r_f-|\mathbf{X}_i-\mathbf{X}_j|),
\end{aligned}
\label{eq:desVarsDef}
\end{equation}
where $N_{r_f}$ is the number of nodes within the filter radius $r_f$; and $|\mathbf{X}_i-\mathbf{X}_j|$ is the Euclidean distance between nodes i and j. Within each hexahedral background element, the $\text{LSF}$ is interpolated tri-linearly. Furthermore, the regularization scheme described below in Section \ref{subsec:optProbForm} is adopted to promote a uniform spatial gradient of the $\text{LSF}$ near the solid-void interface while converging to either a positive or negative target value away from the interface. 

\subsection{Optimization problem formulation} \label{subsec:optProbForm}
\begin{figure}[h!]
	\centering
	\includegraphics[width=1.0\linewidth]{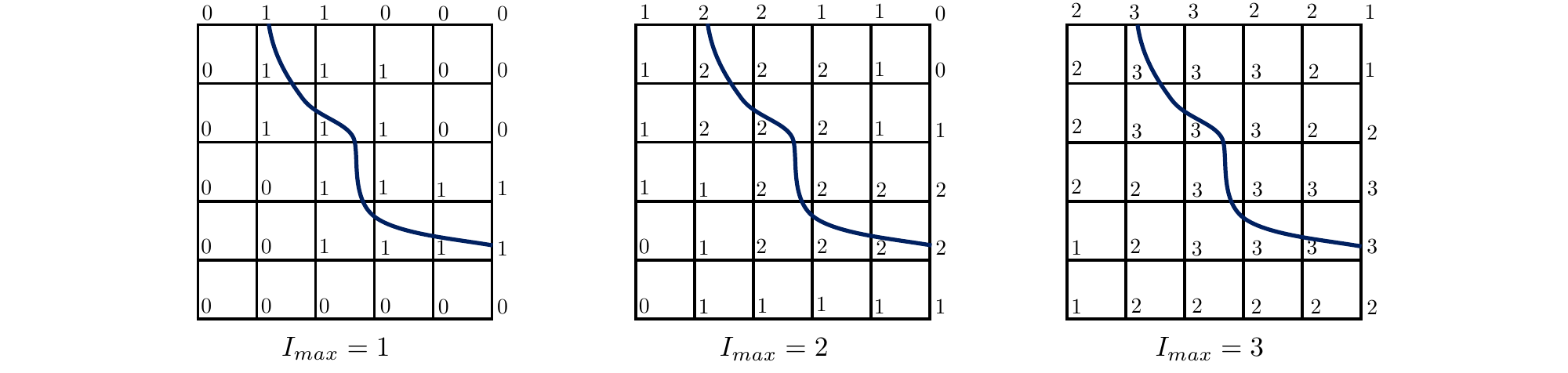}
	\caption{Neighborhood-level field, $I(\mathbf{X})$ , using $I_{max}=[1, 2, 3]$.}
	\label{fig:2DExplanationOfLoN}
\end{figure}

%-----------------------------------------------%
The optimization problems considered in this work are formulated as:
% Equation
\begin{equation}
\begin{aligned}
\underset{\mathbf{s}}{\min}~{z}(\mathbf{s}, \mathbf{u})  = ~
& w_1 {F}(\mathbf{s}, \mathbf{u})  
+ 
w_2 P_{Per}(\mathbf{s})
+
w_3 P_{Reg}(\mathbf{s})
\\
s.t.: ~~~
& g_i(\mathbf{s},\mathbf{u}) \leq 0, i=1,...,N_g.
\\
\mathbf{u} \in  \rm I\!R^{N_\mathbf{u}}, ~&
\mathbf{s} \in \Pi = \{  \rm I\!R^{N_{{s}}}~|~\phi_{low} \leq {s}_i \leq \phi_{up}, i=1,...,N_s \} 
\end{aligned}
\label{eq:optProbSetupForm}
\end{equation}
The first component of the objective, ${z}$, represents the quantity of interest to be minimized, ${F}$, such as strain energy, target displacement, and mass. 
The second term is the normalized perimeter control penalty, added to avoid the emergence of irregular geometric features:
% Equation
\begin{equation}\label{eq:TFCOptProbSetup}
\begin{aligned}
P_{Per}
= 
\frac{\int_{\Gamma_{S,V}}~dA}{\int_{\Gamma_{S,V}^0}~dA},
\end{aligned}
\end{equation}
%t
which is normalized by the perimeter of the initial design, $\Gamma_{S,V}^0$.
The last component is a penalty that aims at regularizing the $\text{LSF}$. 
This penalty ensures (i) an approximately uniform spatial gradient near the solid-void interface; and (ii) that the $\text{LSF}$ away from the interface convergence to target upper and lower bounds.
The normalized formulation of $P_{Reg}$ reads:
% Equation
\begin{equation}\label{eq:LsRegEq}
\begin{aligned}
	P_{Reg} = 
	\frac{1}{\displaystyle\int_{\Gamma_{S,V}^0}~dA} \bigg[ &
\displaystyle\int_{\Omega_D} \alpha_1 (1-w) \left(\displaystyle {\phi(\mathbf{s})}/{\tilde\phi}-sign(\phi) \right)^2 dV + 
                \displaystyle\int_{\Omega_D} \alpha_2 w \left( \displaystyle { || \nabla\phi(\mathbf{s}) || }/{ ||\nabla\tilde\phi|| }-1\right)^2 dV 
\\
&
	+ \displaystyle\int_{\Omega_D} \alpha_3 (1-w) (\left\lVert \nabla\phi(\mathbf{s}) \right\rVert)^2 dV \bigg],
\end{aligned}
\end{equation}
with $\alpha_1$, $\alpha_2$, and $\alpha_3$ as weighting factors.
The first term penalizes the difference between the $\text{LSF}$, $\phi(\mathbf{s})$, and a target value, $\tilde\phi$, away from the interface.
The second term penalizes the difference between the norm of the spatial gradient of the $\text{LSF}$, $\nabla\phi(\mathbf{s})$, and a target gradient norm, $\nabla\tilde\phi$, in the vicinity of the interface. 
The third term enforces a gradient of zero away from the interface.
The parameter $w$ is a measure of the distance of a node to the interface defined as:
% Equation
\begin{equation}\label{eq:LsRegScaling}
\begin{aligned}
	w = 
	e^{\displaystyle -\gamma_w (I(\mathbf{X})/I_{max}-1)^2},
\end{aligned}
\end{equation}
where the parameter $\gamma_w$ determines the proximity of the interface. 
In this regularization scheme, the distance of a node to the interface is approximated by a neighborhood-level field $I(\mathbf{X})$, as described in \cite{Noel2019HB}, and illustrated in Fig.~\ref{fig:2DExplanationOfLoN}.
The $I_{max}$ parameter represents a measure of the of the region along the interface over which the approximation $I(\mathbf{X})$ is evaluated.
The higher $I_{max}$, more background elements surrounding the interface are considered.
In this work, we set $\tilde\phi=3h(sign(\phi))$, $\nabla\tilde\phi=1$, $I_{max}=1$, and the weighting factors $\alpha_i$ in the range of $[0.1-0.5]$.
Furthermore, the weights $w_i$ in the objective function are selected such that the contributions of the regularization and perimeter control penalty components are significantly lower than ${F}$. 

%Figure 

The design needs to satisfy a set of $N_g$ problem dependent inequality constrains, [$g_1$, ..., $g_{N_g}$] that consider limits on, e.g., mass, maximum allowable stress, and minimum eigenvalue.
The number of design variables are denoted by $N_s$ and the number of state variables is $N_u$; see Eq.~\ref{eq:optProbSetupForm}. The state variables $\mathbf{u}$ are governed by a set of discretized partial differential equations, which are satisfied at each design in the optimization process.

The parameter optimization problems are solved using a gradient-based nonlinear programming scheme, namely the Globally Convergent Method of Moving Asymptotes (GCMMA), with no inner iterations; see \cite{svanberg2002class}.

\subsection{Sensitivity analysis}
%-----------------------------------------------%
The design sensitivities of the objective function with respect to the design variables are defined as
% Equation
\begin{equation}\label{eq:SensAnaGenEq}
\begin{aligned}
	\frac{d z}{d \mathbf{s}} = 
        \frac{\partial z}{\partial \mathbf{s}} + 
	 \frac{\partial z}{\partial \mathbf{u}} \frac{d \mathbf{u}}{d \mathbf{s}}.
\end{aligned}
\end{equation}
The first term of the right-hand side represents the explicit dependencies. The second term accounts for the implicit dependencies through the state variables.
The implicit term is evaluated using the adjoint method:
% Equation
\begin{equation}\label{eq:SensAnaAdjEq}
\begin{aligned}
	 \frac{\partial z}{\partial \mathbf{u}} \frac{d \mathbf{u}}{d \mathbf{s}} = 
        - {\boldsymbol{\lambda}}^T 
	 \frac{\partial \mathbf{R}}{\partial \mathbf{s}},
\end{aligned}
\end{equation}
where $\boldsymbol{\lambda}$ is the adjoint response and $\mathbf{R}$ is the residual defined later in Eq.~\ref{eq:resGenForm}. The adjoint linear problem is formulated as:
% Equation
\begin{equation}\label{eq:AdjProb}
\begin{aligned}
	\boldsymbol{\lambda} =
	 \frac{\partial z}{\partial \mathbf{u}} 
	 \left[ \frac{\partial \mathbf{R}}{\partial \mathbf{u}} \right]^{-1}
\end{aligned}
\end{equation}
For more information on the sensitivity analysis within a level set $\text{XFEM}$ topology optimization framework, the reader is referred to \cite{sharma2017shape}.

\subsection{Physical analysis}\label{subsec:XFEMTheory}
%------------------------------%

A generalized Heaviside enriched $\text{XFEM}$ strategy where the degrees of freedom within each unique sub-phase are approximated using standard finite element shape functions is adopted. 
The flexibility of this $\text{XFEM}$ approach has been demonstrated in single and multiphysics topology optimization problems; see for example, \cite{maute2015level,behrou2017level}. 
Here, the response is consistently interpolated in elemental subdomains with the same phase using the formulation shown below.

\subsubsection{XFEM formulation}\label{subsec:XFEMForm}
%------------------------------%
The state variable vector, $\hat{\mathbf{u}}_i(\mathbf{X})$, at node i of a two-material problem (phase I, phase II) is approximated as:
% Equation:
\begin{equation}\label{eq:HeavisideEnrich}
\begin{aligned}
	\hat{\mathbf{u}}_i(\mathbf{X})
	=
	\sum^{L}_{l=1}
	\bigg(
	H(-\phi(\mathbf{X}))\sum^{N_N^e}_{k=1} {\mathcal{N}}_k(\mathbf{X})\delta_{lq}^{k}\mathbf{u}_{il}^{k,\Omega_{S}} 
	+ 
	H(\phi(\mathbf{X}))\sum^{N_N^e}_{k=1} {\mathcal{N}}_k(\mathbf{X})\delta_{lq}^{k}\mathbf{u}_{il}^{k,\Omega_{V}}
	\bigg),
\end{aligned}
\end{equation} 
where $H$ is the Heaviside function depending on the $\text{LSF}$:
% Equation
\begin{equation}\label{eq:HeavisideFunc}
\begin{aligned}
	H(\phi) =
	\begin{cases}
		1, & \forall ~\phi(\mathbf{X}) > 0 \\
		0, & \forall ~\phi(\mathbf{X}) < 0.
	\end{cases}
\end{aligned}
\end{equation}

In Eq.~\ref{eq:HeavisideEnrich}, $L$ denotes the maximum number of enrichment levels, ${\mathcal{N}}_k(\mathbf{X})$ represents the nodal shape function, and $\delta_{lq}^{k}$ is the Kronecker delta which selects the active enrichment level $q$ for node $k$. The Kronecker delta ensures that displacements at node $k$ are only interpolated by a single set of degrees of freedom defined at node position $\mathbf{X}$ such that the partition of unity principle is satisfied (\cite{babuvska1997partition}). The number of nodes per element is given by $N_N^e$. The reader is referred to \cite{makhija2014numerical,terada2003finite,tran2011multiple} for an in-depth explanation of the generalized Heaviside enrichment strategy employed. 

Note that other immersed boundary techniques with a crisp interface definition, such as CutFem  \cite{burman2015cutfem} and HIFEM \cite{soghrati2016application}, are equally applicable without loss of generality. 

\subsubsection{Governing equations}\label{subsec:govEq}
%------------------------------%

Static equilibrium is described by the following residual equation augmented with stabilization terms:
 % Equation:
\begin{equation}\label{eq:resGenForm}
\begin{aligned}
	\mathbf{R}
	=
	\mathbf{R}^U + 
	\mathbf{R}_\Gamma^N  +
	\mathbf{R}_\Gamma^G  +
	\mathbf{R}^S,
\end{aligned}
\end{equation}
in which the weak form of the linear elastic governing equation, $\mathbf{R}^U$, is defined as:
% Equation:
\begin{equation}\label{eq:ResEq}
\begin{aligned}
	\mathbf{R}^U = 	
	\int_{\Omega_{S}} \delta \boldsymbol{\epsilon} : \boldsymbol{\sigma}~ dV - 
	\int_{\Gamma_N^{\Omega_{S}}} \delta \mathbf{u}~\mathbf{T}_\nu~dA.
\end{aligned}
\end{equation}
The displacement field and the test function are denoted by $\mathbf{u}$ and $\delta\mathbf{u}$, respectively. Traction forces, $\mathbf{T}_\nu$, are applied along the boundary $\Gamma_\nu^{\Omega_{S}}$. 
The material tensor for isotropic linear elasticity, $\mathbf{D}$, together with the infinitesimal strain tensor, $\boldsymbol{\epsilon}=\frac{1}{2}(\nabla\mathbf{u} + \nabla\mathbf{u}^T)$, define the Cauchy stress $\boldsymbol{\sigma} = \mathbf{D} :\boldsymbol{\epsilon}$.

The unsymmetric version of Nitsche's method is employed to weakly enforce Dirichlet boundary conditions (\cite{nitsche1971variationsprinzip,burman2012fictitious}). These conditions are applied through the following residual component:
% Equation:
\begin{equation}\label{eq:NitscheFrom}
\begin{aligned}
	\mathbf{R}_\Gamma^N 
	= &
	 - \int_{\Gamma} 
	\llbracket \delta \mathbf{u} \rrbracket ~ \boldsymbol\sigma \cdot \boldsymbol{\nu}~dA 
	+ \int_{\Gamma} 
	\delta ( \boldsymbol\sigma \cdot \boldsymbol{\nu} ) \llbracket \mathbf{u} \rrbracket~dA 
	+ \gamma_N~E / h\int_{\Gamma} 
	\llbracket \delta \mathbf{u} \rrbracket \llbracket \mathbf{u} \rrbracket~dA.
\end{aligned}
\end{equation}
The normal vector on a domain boundary is denoted by $\boldsymbol{\nu}$.
The jump operator $\llbracket \boldsymbol{\cdot} \rrbracket$ is defined as
% Equation:
\begin{equation}\label{eq:jumpOp}
\begin{aligned}
	\llbracket \mathbf{u} \rrbracket = \mathbf{u} - \bar{\mathbf{u}}, ~~~~~
	\llbracket \delta \mathbf{u} \rrbracket = \delta \mathbf{u} - \delta \bar{\mathbf{u}}.
\end{aligned}
\end{equation}
The integrals of Eq.~\ref{eq:NitscheFrom} correspond to the standard consistency, adjoint consistency, and the Nitsche penalty terms, respectively. The third term is scaled by the Young's modulus, $E$, the element size, $h$, and the penalty factor $\gamma_N$. This last term provides additional control over the accuracy at which a boundary condition (BC) is enforced. Our numerical experiments show that choosing a penalty factor within a range of $\gamma_N=[10,100]$ provides an adequate enforcement of BCs while avoiding an ill-conditioned linear system.

To prevent numerical instabilities due to vanishing zones of influence of certain degrees of freedom when the interface approaches a node, face-oriented ghost penalization is used in the vicinity of the interface; see \cite{schott2014new} and \cite{burman2015cutfem}. 
Approximating $\mathbf{u}$ and $\delta \mathbf{u} $ by tri-linear shape functions, ill-conditioning is mitigated by applying the following penalty:
% Equation:
\begin{equation}\label{eq:GhostPenForm}
\begin{aligned}
	\mathbf{R}_\Gamma^G 
	= 
	Eh \gamma_G 
	\underbrace{\sum}_{\mathcal{F} \in \mathcal{F}_{cut}}
	\int_{\mathcal{F}}
	\llbracket \nabla \delta \mathbf{u} \cdot \boldsymbol{\nu_e} \rrbracket 
	\llbracket \nabla \mathbf{u} \cdot \boldsymbol{\nu_e} \rrbracket 
	~dA.
\end{aligned}
\end{equation}
Element faces in the vicinity of the interface for which at least one of the two adjacent elements is intersected are included in $\mathcal{F}_{cut}$, as explained in \cite{villanueva2017cutfem}.
Normals of these element faces are denoted by $\boldsymbol{\nu_e}$.
The influence of the ghost penalty term presented above is controlled by the penalty factor $\gamma_G$, which is usually chosen to be within $\gamma_G=[0.001,0.01]$ based on numerical experiments; see \cite{geiss2018topology, geiss2019regularization}.

To suppress rigid body motion of disconnected solid subdomains that may emerge and develop, selective structural springs are added to the governing equations; see \cite{geiss2018topology,geiss2019combined}. An additional stiffness term is applied only to solid subdomains that are disconnected from the support via the following residual component:
% Equation:
\begin{equation}\label{eq:SelecSprEq}
\begin{aligned}
	\mathbf{R}^S = \gamma_S E / h^2 \int_{\Omega_{D}} \delta \mathbf{u} \cdot \mathbf{u} ~dV.
\end{aligned}
\end{equation}
The parameter $\gamma_S$ denotes the spring stiffness scaling and is non-zero only for the free-floating pieces of solid material. 
These pieces are identified by an auxiliary indicator field constructed from the solution of a diffusion problem, as described in \cite{geiss2019combined}.

In the final example, a gradient stabilized scalar stress field, $\tau(\mathbf{X})$, is post-processed via the $\text{XFEM}$ informed smoothing procedure described in \cite{sharma2018stress}; see Section \ref{ex3:bracket}. The additional set of state variables is computed by solving the residual equation:
% Equation:
\begin{equation}\label{eq:stressProj}
\begin{aligned}
	\mathbf{R}=\mathbf{R}^{\tau}= 
	\int_{\Omega_{I}} ~ \delta \tau ~ (\tau - \sigma_{VM}) ~ dV +
	h^2 \gamma_\tau
	\underbrace{\sum}_{\mathcal{F} \in \mathcal{F}_{cut}}
	\int_{\mathcal{F}}
	\llbracket \delta \nabla \tau \rrbracket 
	\llbracket \nabla \tau \rrbracket 
	~dA.
\end{aligned}
\end{equation}
The field $\sigma_{VM}(\mathbf{X})$ and the parameter $\gamma_\tau$ represent the von Mises stress and the ghost penalty weight, respectively.
Spatial oscillations of $\tau$ are mitigated by penalizing the jump in spatial stress gradients across elemental faces (second term in Eq.~\ref{eq:stressProj}). 

%-------------------------------------------------------------------------------------------------%
\section{Numerical Examples} \label{sec:NumEx}
%-------------------------------------------------------------------------------------------------%

The local and global rules presented in Section \ref{sec:rulesForTopAmb} are studied in this section to assess their influence on the evolution of the topology, and in the optimized design.
Furthermore, the influence of the initial seeding pattern of holes on the optimization process is investigated for different rules.
These rules are tested on structural linear elastic single material, solid-void problems solved by the explicit level set $\text{XFEM}$ topology optimization framework detailed in Section \ref{sec:LsXFEMTopOptFramework}.
A strain energy optimization problem is formulated in Example 1, minimizing the displacement in a portion of the structure is investigated in Example 2, and in the last example, the objective consists of minimizing the total mass while satisfying a stress constraint.
The optimization problem is considered converged if the constraints are satisfied and the relative change in objective between two consecutive design iterations is less than $1\times10^{-3}$. The initial seeding of the design domains with holes is constructed using patterns of cuboid or spherical primitives.

The governing equations are discretized using the $\text{XFEM}$ approach outlined in Section \ref{subsec:XFEMTheory}. 
Relevant parameters used for the numerical examples 1 and 2 are listed in Table \ref{tab:commonOptProbParams} in self-consistent units.
 Table \ref{tab:ex3BracketMatProps} summarizes the parameters used in the last example in SI units.
The example problems consist of a one-way coupled set of governing equations; i.e., the diffusion problem describing the auxiliary indicator field, the stabilized linear elasticity equations (Eq.~\ref{eq:resGenForm}) and, for the third problem, the stabilized stress projection equations (Eq.~\ref{eq:stressProj}).
The systems of equations of the first and second examples are solved using an ILU preconditioner (\cite{saad1996ilum}).
The third example was solved iteratively via a Trilinos algebraic multigrid solver (\cite{heroux2005overview}).

% Table
\begin{table}[h!] 
	\caption{\label{tab:commonOptProbParams}Parameters common to all numerical examples function of the element size $h$.}
	\centering
	\renewcommand{\arraystretch}{1.2}
	\begin{tabular}{l|c}
		\hline
		Parameter                  & Value\\\hline		
Target $\text{LSF}$           		 &  $\tilde\phi = 2.5h$        \\
$\text{LSF}$ regularization control               &   $\gamma_{P_{Reg}} = -log(0.01)$     \\
Filter radius           	 &  $f_r = 1.8h $      \\
Nitsche penalty factor &  $\gamma_N = 100.0 $      \\
Ghost penalty factor	 &  $\gamma_G = 0.01 $      \\
Spring stiffness factor           	 &  $\gamma_S = 1\text{x}10^{-6} $      \\
Stress ghost penalty factor (Ex. 3)          	 &  $\gamma_\tau = 0.01 $      \\
		\hline
	\end{tabular}
\end{table}

%------------------------------------%
\subsection{Beam}
%------------------------------------%

A beam problem is used to investigate the effect of the topology consistency rules detailed in Section \ref{sec:rulesForTopAmb} as the mesh is refined. 
In this example, we focus on analyzing differences in the optimized designs that typically favor thin-walled shear-webs when using a level set topology optimization approach.
Three levels of refinement are considered.

\begin{figure}[h]
	\centering
	\includegraphics[width=1.0\linewidth]{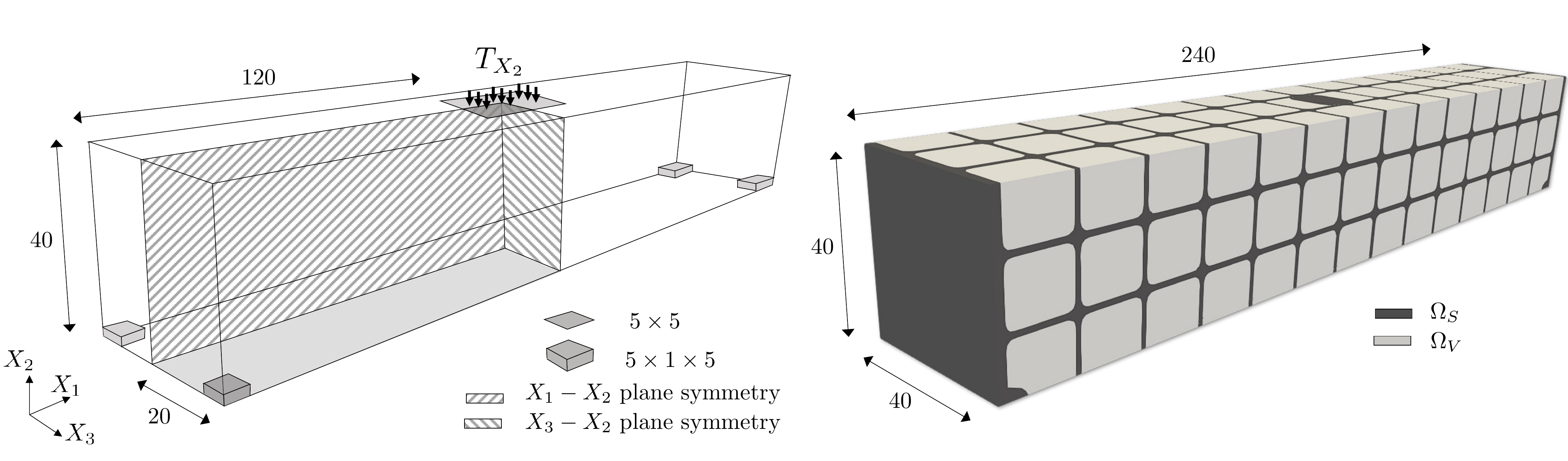}
	\caption{Beam problem: (left) design domain with boundary conditions, symmetry planes and dimensions; and (right) initial design with void inclusions.}
	\label{fig:beamProbSetup}
\end{figure}

The problem setup is shown in Fig.~\ref{fig:beamProbSetup}.
The design domain of size $240 \times 40 \times 40$ is simply supported at all four corners of the bottom face, as shown on the left side of Fig.~\ref{fig:beamProbSetup}.
A traction load $T_{X_2} = -2.0$ is applied on the center region of the top face on a $10\times 10$ area.
Only one quarter of the design domain is simulated by applying symmetry boundary conditions along the $X_1-X_2$ and $X_3-X_2$ planes at the center of the design domain. 
The loading and support regions highlighted in dark gray are excluded from the design domain.
Structured hexahedral meshes with element sizes of $h=[4.0, 2.0, 1.0]$ are considered.
The initial design for this problem is constructed using cuboids as void inclusions.
The right side of Fig.~\ref{fig:beamProbSetup} shows the hole seeding arrangement for the finest mesh used in this example.

%Figure 

The optimization problem seeks to minimize the strain energy, $\Psi$, subject to a mass, $\mathcal{M}$, constraint; and is formulated as follows:
% Equation
\begin{equation}\label{eq:Ex1OptProbSetup}
\begin{aligned}
\underset{\mathbf{s}}{\min}~z(\mathbf{s},\mathbf{u}) & = 
w_1~ \Psi (\mathbf{s},\mathbf{u}) /\Psi_0 
+ 
w_2 ~ P_{Per}(\mathbf{s})
+ 
w_3 ~ P_{Reg}(\mathbf{s})
\\
s.t.: ~~~ g(\mathbf{s})  & = \mathcal{M}(\mathbf{s}) / \mathcal{M}_{T} - \gamma_m \leq 0
\end{aligned}
\end{equation}
The first component of the objective is normalized by the strain energy of the initial design, denoted by $\Psi_0$. The two remainder components are described in Section \ref{subsec:optProbForm}.

In absence of proper hole seeding mechanisms (e.g., using topological derivatives (\cite{sokolowski1999topological,wang2004incorporating,allaire2005structural}) or a density field to nucleate holes in regions of low densities (\cite{barrera2019hole})), the optimization process can converge to sub-optimal designs.
For this reason, continuation schemes are typically used to gradually enforce constraints.
In this example, a mass constraint with $\gamma_m = 0.20$ is enforced using the continuation strategy described in Table \ref{tab:beamProbMassConstContSch} to avoid premature removal of mass resulting in suboptimal designs.
This constraint is reduced every 12 design iterations in a total of 8 steps (i.e., for $\mathcal{D}_{it} \leq 85$).
The mass of the design is normalized by the mass of the entire design domain, $\mathcal{M}_{T}$.
The optimization problem starts feasible for all mesh sizes, i.e., the initial hole seeding satisfies the initial mass constraint.
The weights of the objective are $w_i$ =[ 0.85, 0.05, 0.10]. 

% Table
\begin{table}[h]
	\caption{\label{tab:beamProbMassConstContSch}Continuation parameters of beam problem for $\gamma_m=0.20$ and three mesh sizes.}
	\centering
	\renewcommand{\arraystretch}{1.2}
	\begin{tabular}{l|c|c|c|c|c|c|c|c}
		\hline
		Parameter                  & $\mathcal{D}_{it}=1$ & $\mathcal{D}_{it}=13$ & $\mathcal{D}_{it}=25$ & $\mathcal{D}_{it}=37$ & $\mathcal{D}_{it}=49$ & $\mathcal{D}_{it}=61$ & $\mathcal{D}_{it}=73$ & $\mathcal{D}_{it}=85$  \\\hline
		$\gamma_m|_{h=4.0}$ &  0.66 &  0.58 &  0.50 &  0.42 &  0.35 &  0.30 &  0.25 &  0.20   \\
		$\gamma_m|_{h=2.0}$ &  0.40 &  0.36 &  0.33 &  0.30 &  0.27 &  0.24 &  0.22 &  0.20   \\
		$\gamma_m|_{h=1.0}$ &  0.27 &  0.26 &  0.25 &  0.24 &  0.23 &  0.22 &  0.21 &  0.20   \\				
		\hline
	\end{tabular}
\end{table}

\subsubsection{Effect of ambiguity decider on optimization problem}
%----------------------------%

\begin{figure}[h]
	\centering
	\includegraphics[width=0.95\linewidth]{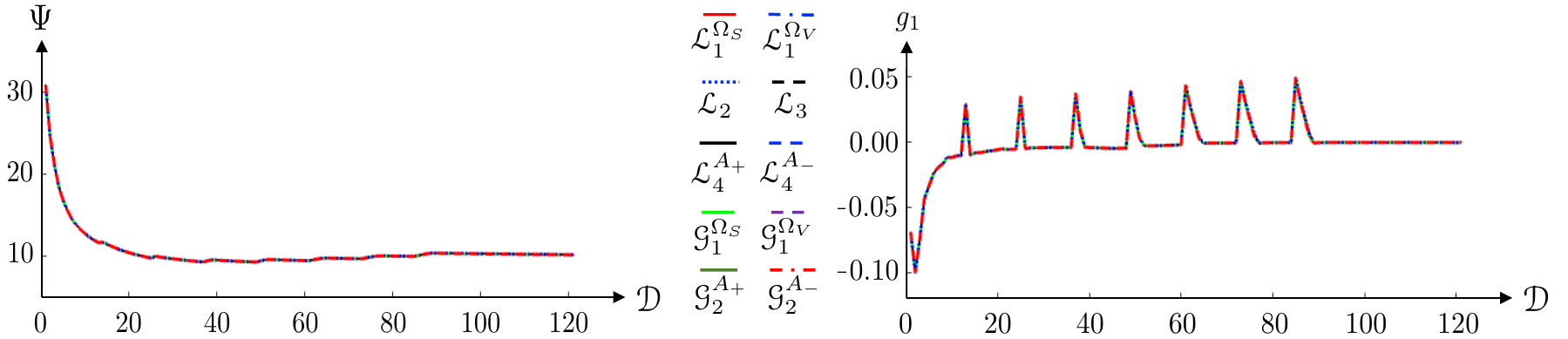}
	\caption{Evolution of (left) objective and (right) mass constraint of beam problem using all ambiguity decider rules.}
	\label{fig:beamEvolObjAndConst}
\end{figure}

The evolution of the strain energy, $\Psi$, and mass constraint, $g_1$, for the results using the finest mesh (i.e., $h=1.0$) are shown in Fig.~\ref{fig:beamEvolObjAndConst}.
The sudden violations of the constraint, and to a lesser extent in the objective, are consequence of the mass constraint being updated at every continuation step. Overall, a smooth behavior is observed in all cases. 
The strain energies of the optimized designs are summarized in Table \ref{tab:FinalStrEnTabBeamProb} and show that the relatively small differences in performance are reduced as mesh is refined.
In this example, the rules used to resolve ambiguities have a negligible effect on both the smoothness of the evolution of the design, as well as the performance of the optimized designs. 

%Figure 

% Table
\begin{table}[h] 
	\caption{\label{tab:FinalStrEnTabBeamProb}Final strain energy of beam problem using three different mesh sizes.}
	\centering
	\renewcommand{\arraystretch}{1.2}
	\begin{tabular}{l|c|c|c|c|c|c|c|c|c|c}
		\hline
Mesh size & $\mathcal{L}_1^{\Omega_{S}}$ & $\mathcal{L}_1^{\Omega_{V}}$ & $\mathcal{L}_2$ & $\mathcal{L}_3$ & $\mathcal{L}_4^{A_{+}}$ & $\mathcal{L}_4^{A_{-}}$ & $\mathcal{G}_1^{\Omega_{S}}$ & $\mathcal{G}_1^{\Omega_{V}}$ & $\mathcal{G}_2^{A_{+}}$ & $\mathcal{G}_2^{A_{-}}$ \\ \hline		
$h = 4$ & $9.0674$ & $8.9884$ & $9.0244$ & $8.9847$ & $9.0749$ & $8.9856$ & $8.9671$ & $9.0354$ & $9.0574$ & $8.9934$  \\
$h = 2$ & $9.1503$ & $9.1359$ & $9.1342$ & $9.1142$ & $9.0998$ & $9.1377$ & $9.1357$ & $9.1537$ & $9.1323$ & $9.1301$  \\
$h = 1$ & $9.9150$ & $9.9133$ & $9.9140$ & $9.9114$ & $9.9134$ & $9.9134$ & $9.9111$ & $9.9121$ & $9.9103$ & $9.9137$  \\
		\hline
	\end{tabular}
\end{table}

%Figure  
\begin{figure}[h!]
	\centering
	\includegraphics[width=0.95\linewidth]{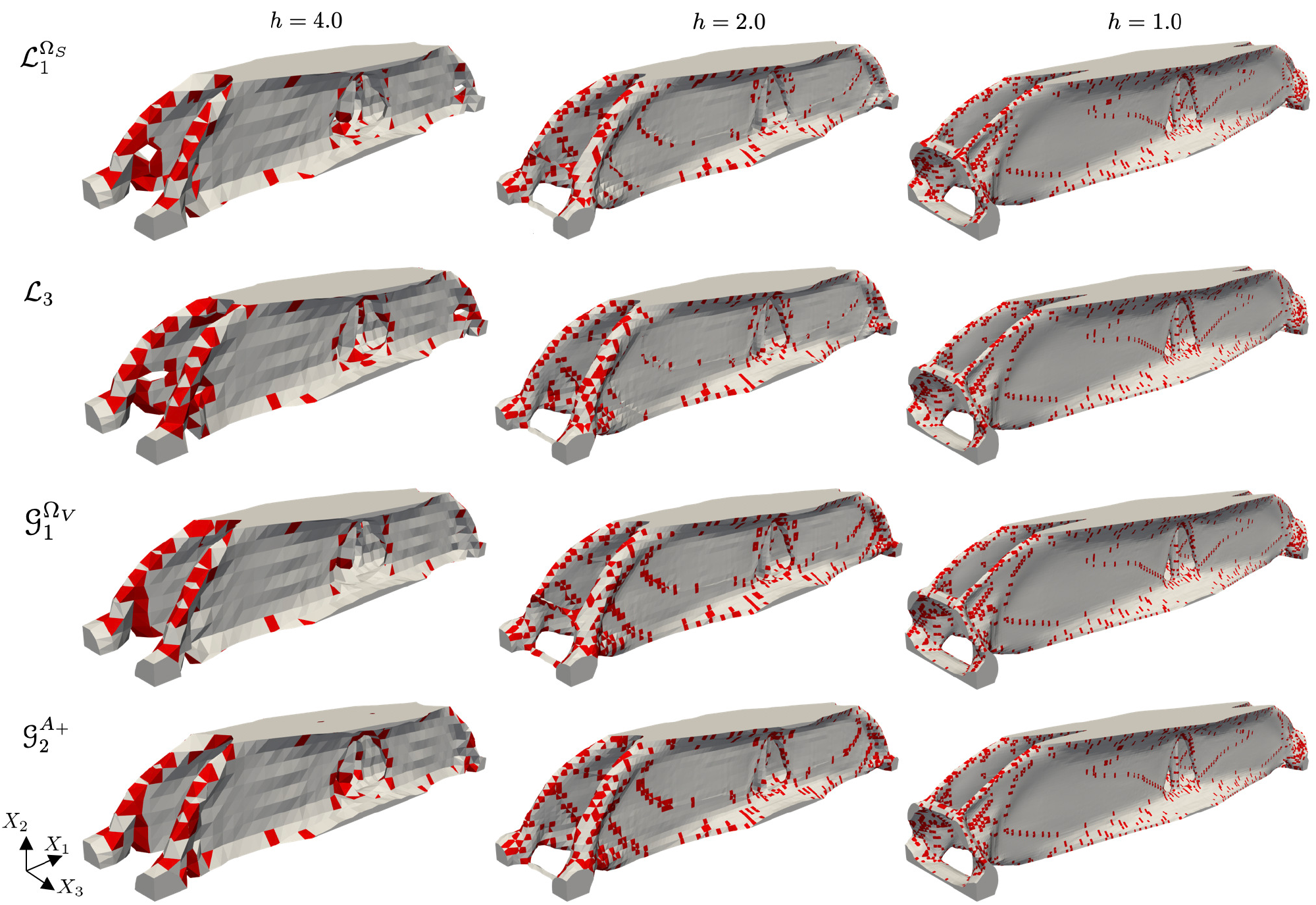}
	\caption{Optimized designs of selected rules using meshes with element size $h=[4.0,2.0,1.0]$.}
	\label{fig:beamFinalDesignsAmbCasesSeparated}
\end{figure}

The optimized designs shown in Fig.~\ref{fig:beamFinalDesignsAmbCasesSeparated} for all three refinement levels and $\mathcal{L}_1^{\Omega_{S}}$, $\mathcal{L}_3$,  $\mathcal{G}_1^{\Omega_{V}}$, and $\mathcal{G}_2^{A_{+}}$ rules show that differences in final topologies are imperceptible in this example.
Although not shown, almost identical designs where obtained using the remaining rules.
Continuous walls connecting the two flanges at the top and bottom of the beam with a thickness of about the element edge length are observed.
Ambiguous tetrahedra, highlighted in red, are observed throughout the optimization process and stem exclusively from hexahedra with one intersection. 
Regardless of the topology consistency rule employed, the thickness of the shear webs in the final structures is slightly larger than the element lengths. 
Hence, holes in such walls that may be consequence of resolving ambiguities using some of the rules described in this paper are not observed in the optimized designs.
The lack of ambiguities that can create connections throughout the evolution of the design explains the little variations in the optimized designs.

%Figure 
\begin{figure}[h!]
	\centering
	\includegraphics[width=0.95\linewidth]{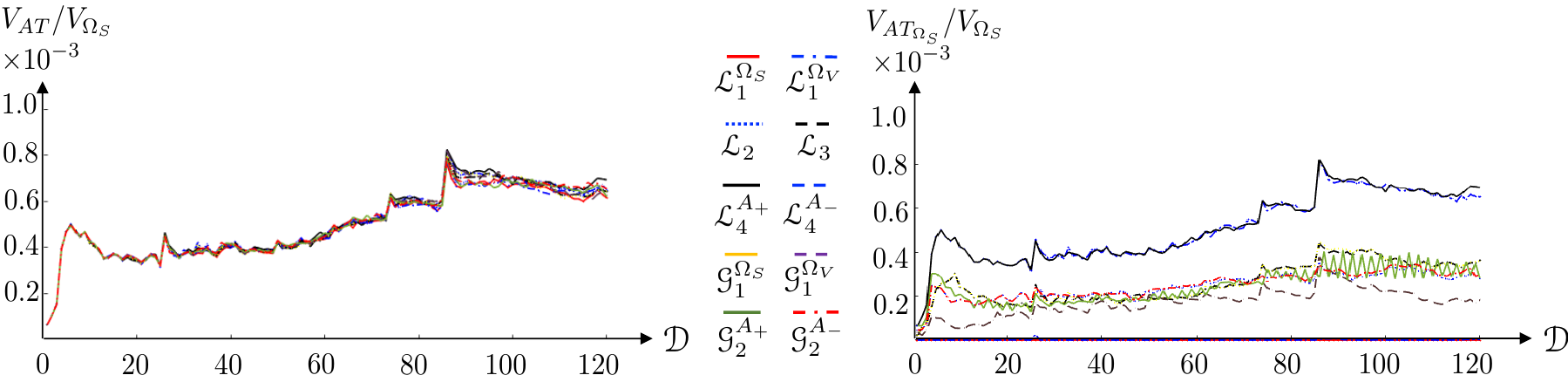}
	\caption{Evolution of ambiguous elements throughout the beam optimization problem: (left) ratio between volume of all $\text{ATs}$ and total volume of solid phase; and (right) ratio between volume of $\text{ATs}$ defined as solid and total volume of solid phase.}
	\label{fig:beamEvolOPenTetsInMesh}
\end{figure}

The evolution of the ratio of $\text{ATs}$ with respect to the total solid volume, $V_{AT} / V_{\Omega_{S}}$, is plotted on the left side of Fig.~\ref{fig:beamEvolOPenTetsInMesh}. 
A slight increase in the disparity between rules of these ratios is observed as the optimization problem converges.
Nevertheless, the $\text{AT}$ volume fraction remains under $0.1\%$ in all cases.
The plot on the right of Fig.~\ref{fig:beamEvolOPenTetsInMesh} shows the ratio of the volume of $\text{ATs}$ for which a solid phase was assigned with respect to the total solid volume, $V_{AT_{\Omega_S}} / V_{\Omega_{S}}$.
As expected, global rules $\mathcal{G}_1^{\Omega_{S}}$ and $\mathcal{G}_1^{\Omega_{V}}$ provide upper and lower limits for the cases considered; see Section \ref{subsec:gblFxdPhaseAssigned}.
Furthermore, it can be seen that rules $\mathcal{L}_2$, $\mathcal{L}_3$, $\mathcal{L}_4$, and $\mathcal{G}_2$ tend to distribute the phase assignment more evenly between the void and solid subdomains.
In all cases, the volume of the $\text{ATs}$ is, at all times, considerably smaller than the uniquely defined domain.

\subsubsection{Shear web structures and mass constraint}

To demonstrate that the avoidance of ambiguities along the shear webs are not specific for the particular mass constraint studied above, lower target mass constraints, i.e., $\gamma_m = [0.10,0.15]$, are examined next. 
The only difference with respect to the previous problem setup are the continuation parameters; see Table \ref{tab:beamProbMassConstContSch_lowerMassConst}.
The analysis for this modified setup focuses on rules $\mathcal{G}_1^{\Omega_{S}}$ and $\mathcal{G}_1^{\Omega_{V}}$ and a mesh size of $h=1.0$ since they provide the widest variations in the problem's geometry at any particular design iteration.

% Table
\begin{table}[h]
	\caption{\label{tab:beamProbMassConstContSch_lowerMassConst}Updates of mass constraint of beam problem for $\gamma_m=[0.10,0.15]$.}
	\centering
	\renewcommand{\arraystretch}{1.2}
	\begin{tabular}{l|c|c|c|c|c|c|c|c}
		\hline
		Parameter                  & $\mathcal{D}_{it}=1$ & $\mathcal{D}_{it}=13$ & $\mathcal{D}_{it}=25$ & $\mathcal{D}_{it}=37$ & $\mathcal{D}_{it}=49$ & $\mathcal{D}_{it}=61$ & $\mathcal{D}_{it}=73$ & $\mathcal{D}_{it}=85$  \\\hline
		$\gamma_m|_{h=1.0}$ &  0.27 &  0.25 &  0.23 &  0.21 &  0.19 &  0.17 &  0.16 &  0.15   \\
		$\gamma_m|_{h=1.0}$ &  0.27 &  0.24 &  0.21 &  0.18 &  0.16 &  0.14 &  0.12 &  0.10   \\				
		\hline
	\end{tabular}
\end{table}

\begin{figure}[]
	\centering
	\includegraphics[width=0.95\linewidth]{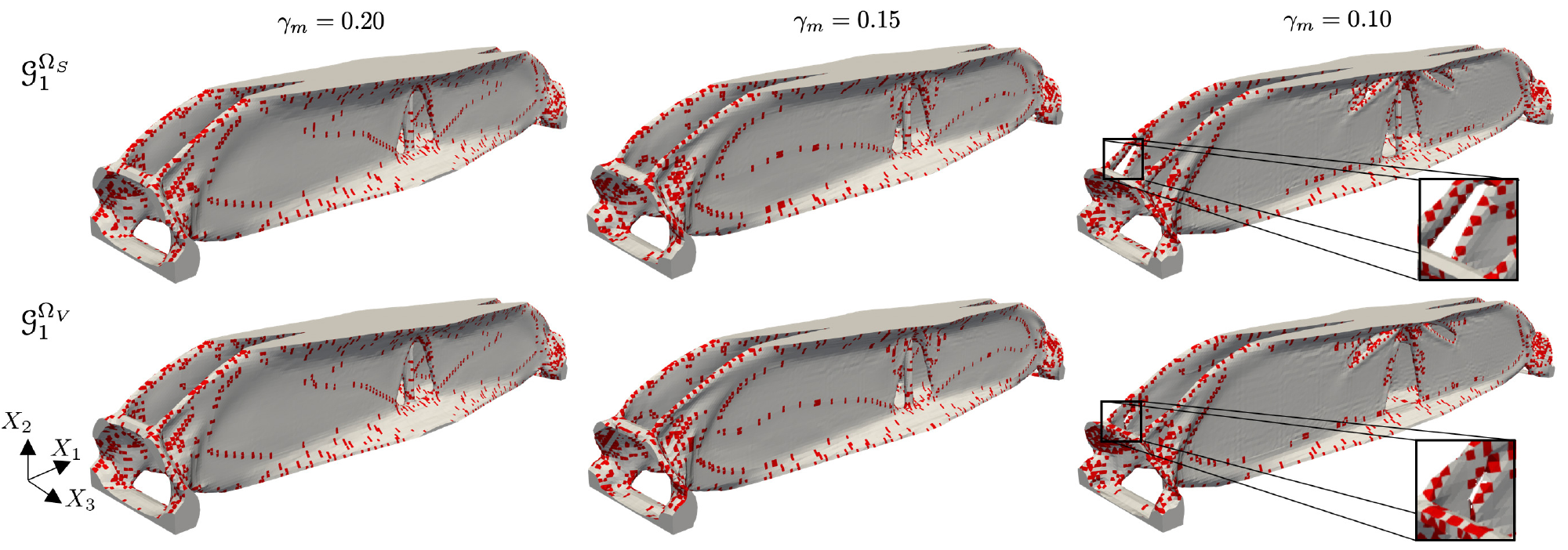}
	\caption{Optimized designs of beam problem using the $\mathcal{G}_1^{\Omega_{S}}$ and $\mathcal{G}_1^{\Omega_{V}}$ rules, and for $\gamma_m = [0.10,0.15,0.20]$.}
	\label{fig:beamFinalDesignsDiffMassConstraint}
\end{figure}

Figure~\ref{fig:beamFinalDesignsDiffMassConstraint} shows the optimized designs using the $\mathcal{G}_1^{\Omega_{S}}$ and $\mathcal{G}_1^{\Omega_{V}}$ phase assignment rules, for decreasing mass constraint from left to right and highlighting $\text{ATs}$ in red. 
It can be seen that the reduction in mass requirements is satisfied by generating new holes rather than decreasing the thickness of the vertical thin walls.
A small influence on the topology consistency rule employed is observed on the shape and size of these holes; see onsets of right column of Fig.~\ref{fig:beamFinalDesignsDiffMassConstraint}. 
In this example, all $\text{ATs}$ are internal, i.e., the configuration shown in Fig.~\ref{fig:hexesWithAmbSubElemsBothCases}(b).
Hence, determine ambiguities as solid or void in the optimized designs can only generate shape changes while keeping the same topology.
However, optimized designs can also include $\text{BATs}$ resulting from hexahedra with multiple intersections. This is shown in Example 3 using a different objective formulation.

%Figure  

%------------------------------------%
\subsection{Structural suspender}
%------------------------------------%

\begin{figure}[h]
	\centering
	\includegraphics[width=1.0\linewidth]{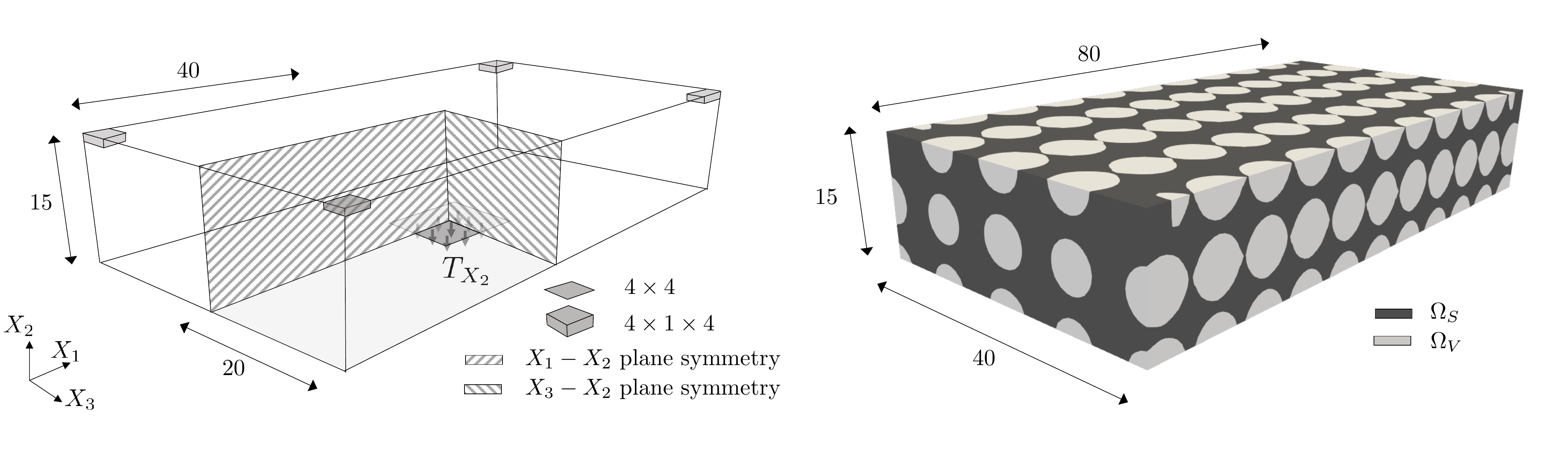}
	\caption{Structural suspender problem: (left) design domain with boundary conditions, symmetry planes and dimensions; and (right) hole seeding arrangement in initial design.}
	\label{fig:suspenderProbSetup}
\end{figure}

In the second problem, we investigate the evolution of a design when the initial topology is significantly affected by the rule chosen to resolve ambiguities.
The design domain of size $80\times 40\times 15$ is subject to the boundary conditions specified on the left side of Fig.~\ref{fig:suspenderProbSetup}.
All four top corners of the domain are clamped in all three directions within the volumes highlighted in Fig.~\ref{fig:suspenderProbSetup}.
A prescribed traction of $T_{X_2} = -10.0$ is applied in the vertical direction in a $4\times4$ rectangular region at the center of the bottom face. 
Symmetry conditions along the $X_1-X_2$ and $X_3-X_2$ planes at the center of the domain are considered. 
The domain is discretized using a mesh with an element edge length of $h=0.71$. 

%Figure 

The design problem in this example consists of minimizing the displacement at a center region, which is identical to the loaded region, at the bottom of the domain. The optimization problem, which resembles the compliant minimization problem studied in the previous example, reads:
% Equation
\begin{equation}\label{eq:Ex2OptProbSetup}
\begin{aligned}
\underset{s}{\min}~z(\mathbf{s},\mathbf{u})  & = 
w_1~ {F}^{U_{X_2}}(\mathbf{s},\mathbf{u}) / {F}^{U_{X_2}}_0 
+ 
w_2 ~ P_{Per}(\mathbf{s})
+ 
w_3 ~ P_{Reg}(\mathbf{s})
\\
s.t.: ~~~ g(\mathbf{s}) & = \frac{\mathcal{M}(\mathbf{s})}{\mathcal{M}_T} - \gamma_m \leq 0,
\end{aligned}
\end{equation}
where ${F}^{U_{X_2}}$ is defined as:
% Equation:
\begin{equation}\label{eq:dispUyForm}
\begin{aligned}
	{F}^{U_{X_2}} = 
        \int_{\Gamma_{U_{X_2}}} U_{X_2}^2~dA,
\end{aligned}
\end{equation}
and represents the average displacement in the $X_2$ direction on a surface $\Gamma_{U_{X_2}}$, which coincides with the surface on which the traction load $T_{X_2}$ is applied.
The remainder second and third components of the objective function are explained in Section \ref{subsec:optProbForm}.
The objective weights are $w_i$ =[0.90, 0.05, 0.05], and a mass constraint with $\gamma_m = 0.15$ is enforced through a continuation scheme. The continuation step size is 12, and the maximum mass allowed decreases by setting $\gamma_m = [0.53, 0.46, 0.40, 0.35, 0.30, 0.25, 0.21, 0.18, 0.15]$.
Both the supports and loading subdomains are excluded from the design domain.

%Figure 
\begin{figure}[b]
	\centering
	\includegraphics[width=1.0\linewidth]{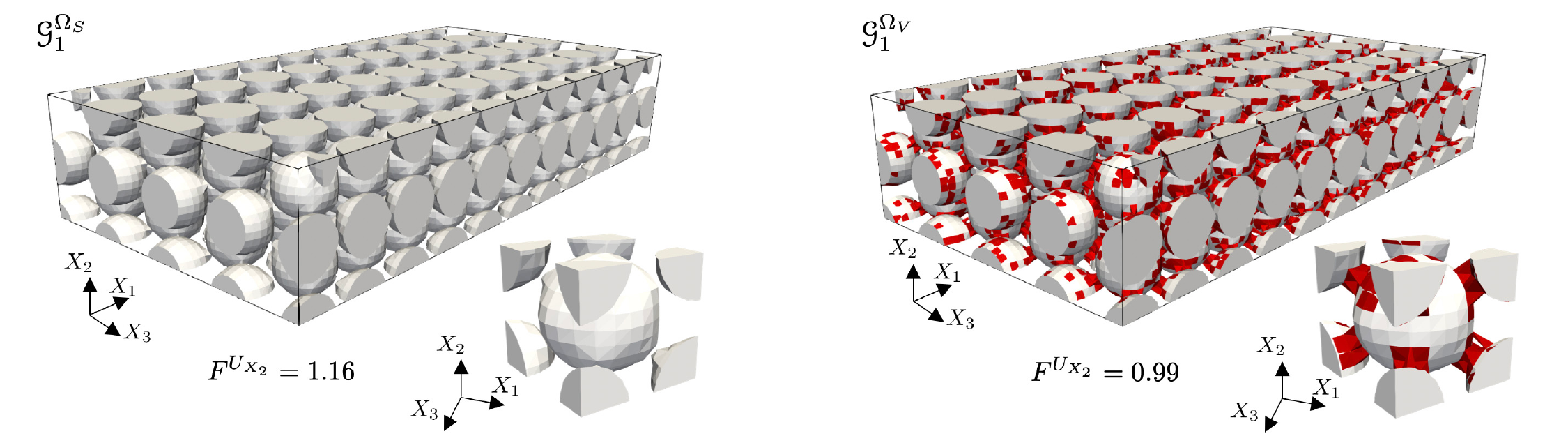}
	\caption{Difference in the initial topology of the structural suspender problem (void phase) as a result of the phase assignment rule used.}
	\label{fig:suspenderInitialSeeding}
\end{figure}

The initial seeding for this problem consists of spherical holes placed such that elements of the structured hexahedral mesh have multiple intersections, as shown on the right side of Fig.~\ref{fig:suspenderProbSetup}. 
In this scenario, both internal and external ambiguities are present and connections between the seeded holes may occur. 
The analysis of this problem focuses on rules $\mathcal{G}_1^{\Omega_{S}}$ and $\mathcal{G}_1^{\Omega_{V}}$ since these rules provide the greatest difference in the resulting geometries.
To better visualize this, the additional volume defined as void when using the $\mathcal{G}_1^{\Omega_{V}}$ rule is highlighted in red on the right side of Fig.~\ref{fig:suspenderInitialSeeding}. 
The void subdomain is shown in this case. 
It can be seen that rule $\mathcal{G}_1^{\Omega_{V}}$ generates connections between the void spheres.
These connections reduce the solid volume by $\approx$3\%. 
Also, note that there is approximately a15\% difference between the initial average displacements.

\begin{figure}[h]
	\centering
	\includegraphics[width=1.0\linewidth]{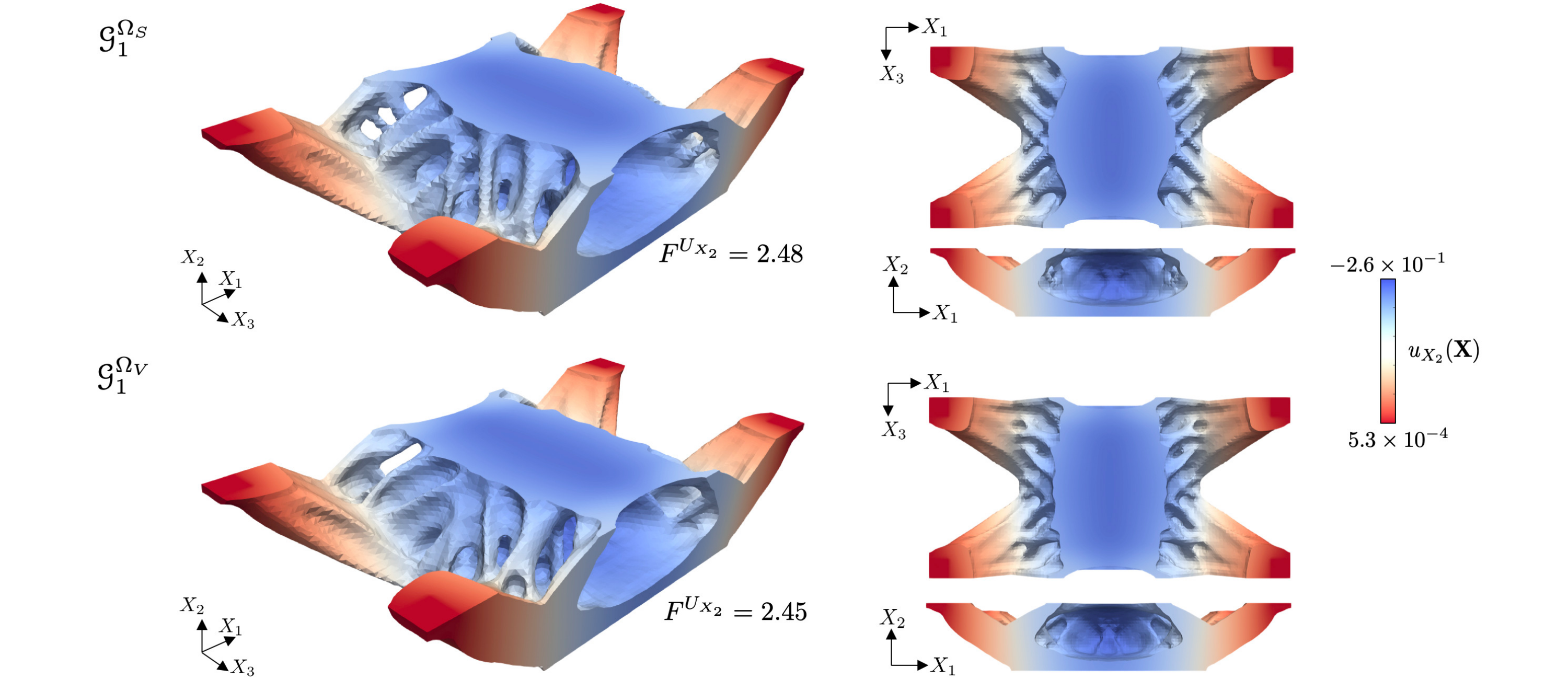}
	\caption{Optimized designs of suspender problem using the $\mathcal{G}_1$ global rules.}
	\label{fig:suspenderIniAndFinalDes}
\end{figure}

Figure \ref{fig:suspenderIniAndFinalDes} shows a 3D view and two plane views of the optimized designs obtained using the $\mathcal{G}_1^{\Omega_{S}}$ and $\mathcal{G}_1^{\Omega_{V}}$ rules on the top and bottom halves, respectively. 
Both designs converge to four legs spreading from the top corner clamped subdomains to the region at the bottom where the distributed load is applied.
Different arrangements of truss-like connections between the legs and the center of the domain on the top are observed.
Overall, the optimized structures differ only locally, and similar performances are achieved despite starting from significantly different initial objectives. 
Hence, in this example, the optimization framework compensates for the phase assignment rule and develop similar discretized geometries.

%Figure 

\begin{figure}[h!]
	\centering
	\includegraphics[width=1.0\linewidth]{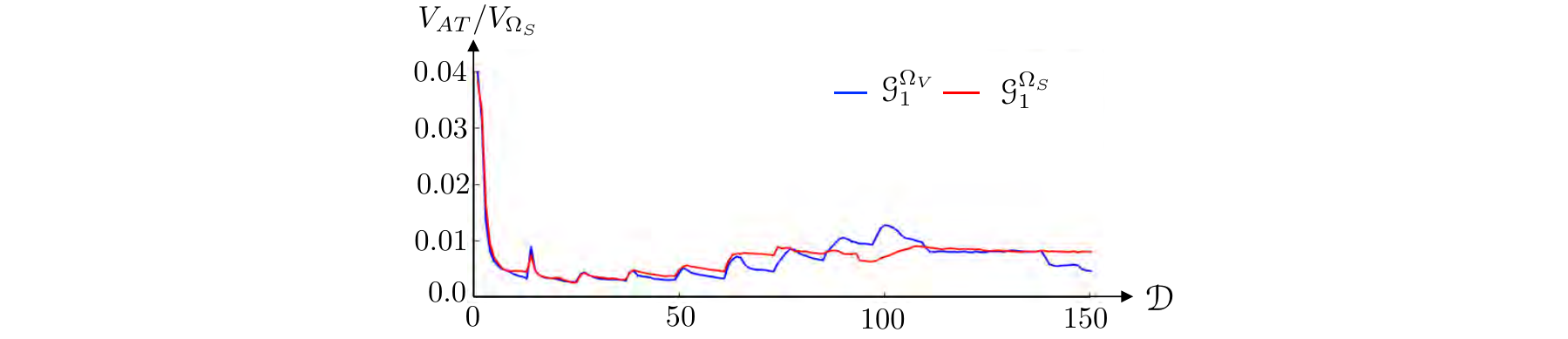}
	\caption{Evolution of ratio between volume of all $\text{ATs}$ and total volume of solid phase throughout the structural suspender optimization problem.}
	\label{fig:suspenderEvolOpenTets}
\end{figure}

The evolution of the $V_{AT} / V_{\Omega_{S}}$ ratio is plotted in Fig.~\ref{fig:suspenderEvolOpenTets}. 
Note the drastic reduction in the total volume of $\text{ATs}$ at the initial stages of the optimization process. 
The $\text{ATs}$ in the optimized designs represent approximately $1\%$ in the extreme cases examined. 
Similar to the previous example, the majority of these ambiguities are originated from intersected hexahedra with a single interface.
However, unlike in the previous example, here the $\text{ATs}$ on the surface produce more pronounced differences in smoothness.

%Figure 

Note that in problems with a strict mass requirement such as the one shown here, a large number of holes may need to be seeded relatively close to each other since they have to cover a large portion of the design domain. 
In such scenarios, rules that favor void phase assignment, i.e. $\mathcal{L}_1^{\Omega_{V}}$ and $\mathcal{G}_1^{\Omega_{V}}$, can make the initial design unfeasible if the relative difference in volume of $\text{ATs}$ is significant.
For this reason, using such rules can result in premature merging of holes and convergence to sub-optimal designs. 

%----------------------------%
\subsection{Bracket} \label{ex3:bracket}
%----------------------------%

In the final example, all local and global rules described in this paper are studied with a realistic engineering design problem.
This problem is characterized by a non-standard 3D geometry of the design domain, and complexities associated with multiple objective components and a stress constraint.
The objective of this optimization problem consists of finding a structure that supports a payload given a set of supports and bolts for attaching the structure to the payload.

\begin{figure}[h]
	\centering
	\includegraphics[width=1.0\linewidth]{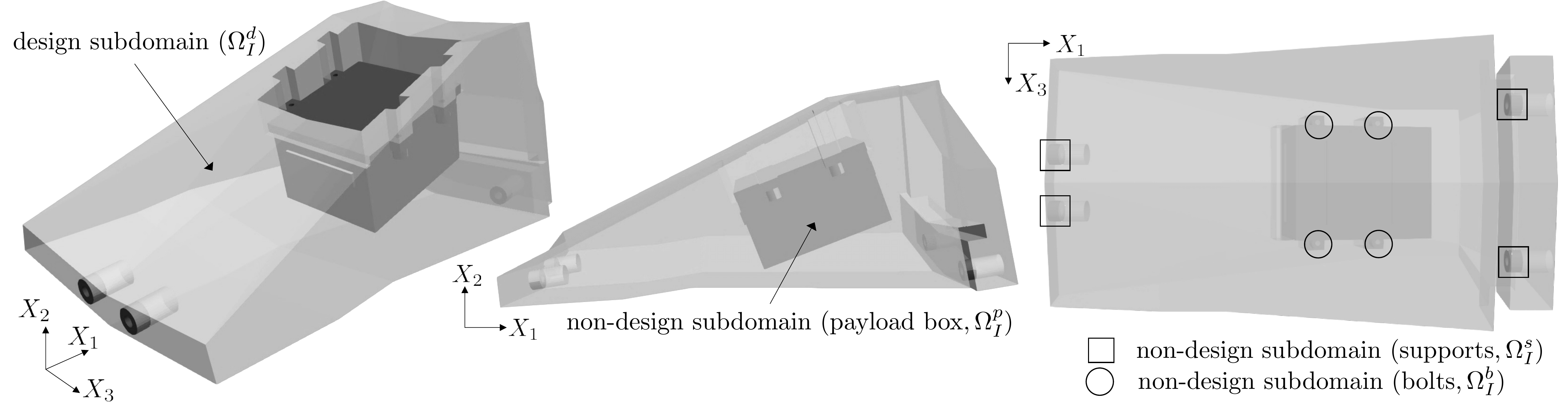}
	\caption{Bracket problem domain highlighting: (left) payload design subdomain; and (center, right) supports and bolts non-design subdomains.}
	\label{fig:bracket3DProbSetup}
\end{figure}

The problem design and non-design subdomains are illustrated in Fig.~\ref{fig:bracket3DProbSetup}.
The design subdomain ($\Omega_{I}^{d}$) is colored in light gray, while the non-design subdomains for the payload box ($\Omega_{II}^{p}$), the supports ($\Omega_{II}^{s}$), and bolts ($\Omega_{II}^{b}$) are colored in dark gray. 
A 3D view together with side and bottom views are provided for a better visualization of the constrained subdomains.  
A uniform pressure load acts on the top surface of the payload box. 
The entire structure is subject to a body force in $X_2$ direction, representing an equivalent shock loading. 
The material properties and load conditions for this static problem can be found in Table \ref{tab:ex3BracketMatProps}.

%Figure 

% Table
\begin{table}[h]
	\caption{\label{tab:ex3BracketMatProps}Material properties and load conditions for bracket problem.}
	\centering
	\renewcommand{\arraystretch}{1.2}
	\begin{tabular}{l|c}
		\hline
		Property                  & Value\\\hline
		Young's Modulus ($\Omega_{S}$, solid)  	 &  $E_S = 1.138$x$10^7$  [N/$cm^2$]\\
		Young's Modulus ($\Omega_{V}$, void)           &  $E_V = 1.138$x$10^{-1}$ [N/$cm^2$] \\
		Material Density ($\Omega_{S}$, solid)     	 		 &  $\theta_S = 4.43$x$10^{-5}$ [kg/$cm^3$]\\
		Material Density ($\Omega_{V}$, void)          		 &  $\theta_V = 0.0$ [kg/$cm^3$]\\
		Poisson Ratio ($\Omega_{S}$ and $\Omega_{V}$) &  $\nu_S = \nu_V$ = 0.342 [-] \\
		Pressure load ($T_{X_2}$) &  1.2x$10^4$ [N/$cm^2$] \\	
		Maximum stress ($\boldsymbol{\sigma}_{max}$) &  398.7 [N/$cm^2$] \\				
		\hline
	\end{tabular}
\end{table}

The goal of this optimization problem is to minimize the mass, $\mathcal{M}$, of the structure that supports the payload box.
The optimization problem is formulated as follows:
% Equation
\begin{equation}\label{eq:Ex3OptProbSetup}
\begin{aligned}
\underset{s}{\min}~{z}(\mathbf{s},\mathbf{u}) & = 
z_{sc} \bigg(
w_1~\mathcal{M} (\mathbf{s})  
+ 
w_2~\Psi (\mathbf{s},\mathbf{u})/\Psi_0
+ 
w_3~P_{\hat\phi}(\mathbf{s})
+ 
w_4~P_{Per}(\mathbf{s})
+
w_5~ P_{Reg}(\mathbf{s}) 
\bigg)
\\
s.t.: ~~~
g_1(\mathbf{s}) & = \mathcal{M} (\mathbf{s}) /\mathcal{M}_{T} - \gamma_m\leq 0; ~~~~~
g_2(\mathbf{s}) = w_{P_{\hat\phi}} ~ P_{\hat\phi}(\mathbf{s}) \leq 0; ~~~~~
g_3(\mathbf{s},\mathbf{u}) = w_{P_{\tau}} ~  P_{\tau}({\mathbf{s},\mathbf{u}}) \leq 0.
\end{aligned}
\end{equation}

The first term in the objective is the mass of the structure in the design subdomain. 
The second term is a strain energy component added to promote a sufficiently stiff structure and prevent an overly aggressive removal of mass early in the design process.
The penalty term, $P_{\hat\phi}$, in $z$ and $g_2$ enforces constraints on the geometry and is formulated as follows:
% Equation
\begin{equation}\label{eq:LsBracketNonDomPen}
\begin{aligned}
P_{\hat\phi} & = 
\frac{\displaystyle\int_{\Omega_D} ( \phi(\mathbf{X}) - \hat\phi(\mathbf{X}) )^2 dV }{\displaystyle\int_{\Gamma_D}~dA},
\end{aligned}
\end{equation}
with the field $\hat\phi(\mathbf{X})$ defined as:
% Equation
\begin{equation}\label{eq:tildeLSFuncBracket}
\begin{aligned}
\hat\phi(\mathbf{X}) & = 
\begin{cases}
\phi_{low}, & \forall~\boldsymbol X \in \Omega_D / (\Omega^{d}_{I} \cup \Omega^{p}_{I} \cup \Omega^{s}_{I} \cup \Omega^{b}_{I}), \\
\phi_{low}, & \forall~\boldsymbol X \in \Gamma_{(\Omega^{d}_{I} \cap \Omega^{p}_{I}) / \Omega^{b}_{I}  }, \\
\phi_{up}, & \forall~\boldsymbol X \in (\Omega^{p}_{I} \cup \Omega^{s}_{I} \cup \Omega^{b}_{I}), \\
\phi(\mathbf{X}), & otherwise;
\end{cases}
\end{aligned}
\end{equation}
based on the previously defined subdomains highlighted in Fig.~\ref{fig:bracket3DProbSetup}.
This penalty (i) forces the support structure to stay within the design domain, (ii) prevents contact between the payload box and the support structure except for the bolts, and (iii) avoids altering the $\text{LSF}$ in the payload box, attachment bolts and the supports.
The remaining penalty terms in the objective ($P_{Per}$ and $P_{Reg}$) are defined in Section \ref{subsec:optProbForm}.
The weights employed in this problem in the objective components and constraints are $w_1=5.0$, $w2 = 0.005$, $w_3=w_{P_{\hat\phi}}=5000.0$, $w_4=0.0001$, $w_5=0.01$, and $w_{P_{\tau}}=1.0\times10^4$.
The objective scaling parameter is set to $z_{sc} = 10.0$. 

A mass constraint ($g_1$, with $\gamma_m=0.30$) imposes an upper limit on the mass, in addition to considering the mass in the objective.
The stress constraint, $g_3$, controls the stress field such that is does not exceed $\sigma_{VM}^{max}$ through the penalty:
% Equation
\begin{equation}\label{eq:ExStressConstForm}
\begin{aligned}
P_{\tau}
=
\int_{\Omega_{d}} \Upsilon dV.
\end{aligned}
\end{equation}
This penalty term measures the volume where the von Mises stress exceeds the stress limit. To this end, the integrand, $\Upsilon$, is increased as the stress exceeds the stress limit. To ensure differentiability at $\Upsilon = \sigma_{VM}^{max}$, the integrand is formulated as follows: 
% Equation
\begin{equation}\label{eq:ExStressFieldForm}
\begin{aligned}
\Upsilon 
=
\begin{cases}
\left[  ((\tau - \sigma_{VM}^{max})/\sigma_{VM}^{max})^2 + \xi^2_{\Upsilon} \right]^{1/2} - \xi_{\Upsilon}, & \forall~\tau- \sigma_{VM}^{max}>0,\\
0, & \forall~\tau- \sigma_{VM}^{max} \leq 0,
\end{cases} 
\end{aligned}
\end{equation}
where the parameter $\xi_\Upsilon >0$ controls the smoothness of $\Upsilon$ for $\tau > \sigma_{VM}^{max}$, and is set to 0.1. The constant $\sigma_{VM}^{max}$ is given by the yield stress of Ti-6Al-4V, reduced by a safety factor of 2.0; see Table \ref{tab:ex3BracketMatProps}.
Note that, although initially inactive, this stress constraint regulates the mass removal effect of the objective at the final stages of the optimization process.

\begin{figure}[h]
	\centering
	\includegraphics[width=1.0\linewidth]{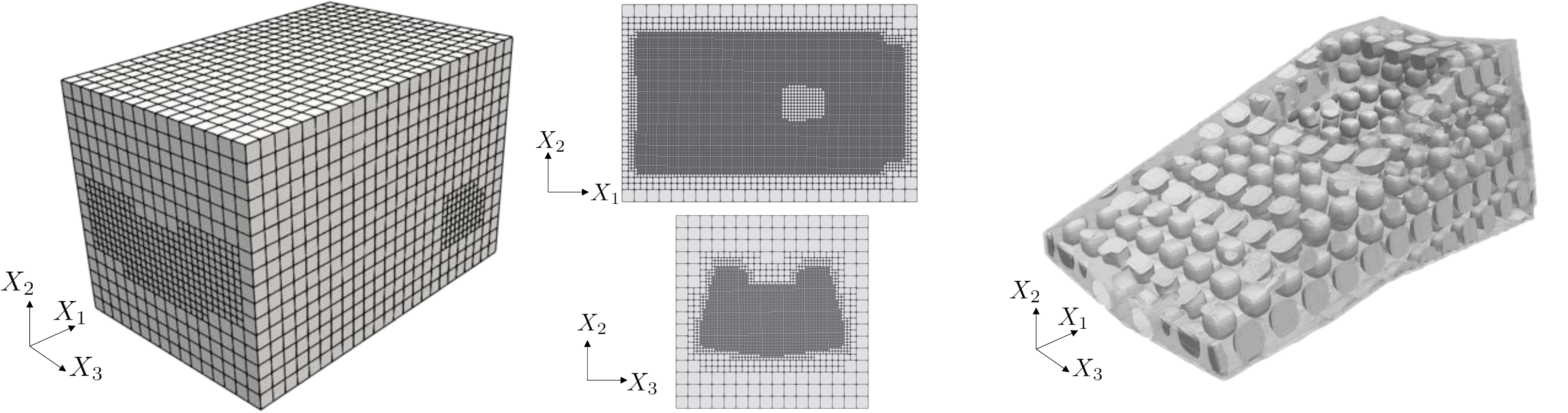}
	\caption{Bracket problem analysis setup: (left) 3D view of locally refined mesh used together with two plane views; and (right) initial seeding in design domain.}
	\label{fig:bracketMeshViews}
\end{figure}

The design domain, supports and bolts are immersed into a hexahedral computational domain of size $15.24\times10.16\times10.16$ $cm^3$. The computational domain is discretized by a hierarchically refined tensor mesh. This mesh is locally refined within the design domain to reduce computational cost while keeping a sufficient resolution in relevant regions.
A 3D view and two plane views at the center in the $X_1-X_2$ and $X_2-X_3$ planes are shown on the left side and center of Fig.~\ref{fig:bracketMeshViews}, respectively.
The initial hole seeding pattern is shown on the right side of Fig.~\ref{fig:bracketMeshViews}.
Note that, unlike in Example 2, in this setup no connections are created between void inclusions with any of the rules employed.

%Figure 

The mass and strain energy of the optimized designs are summarized in Table \ref{tab:FinalObjBracketProb}.
The maximum difference in the mass of the optimized designs across all design rules is less than 0.3\%.
The differences in strain energy are negligible as well.
Similar to what was observed in the previous examples, the performance is not compromised by any of the topology consistency rules tested.

% Table
\begin{table}[h!] 
	\caption{\label{tab:FinalObjBracketProb}Final mass and strain energy of bracket example using all local and global rules.}
	\centering
	\renewcommand{\arraystretch}{1.2}
	\begin{tabular}{l|c|c|c|c|c|c|c|c|c|c}
		\hline
$\mathcal{Z}$ component & $\mathcal{L}_1^{\Omega_{S}}$ & $\mathcal{L}_1^{\Omega_{V}}$ & $\mathcal{L}_2$ & $\mathcal{G}_3$ & $\mathcal{L}_4^{A_{+}}$ & $\mathcal{L}_4^{A_{-}}$ & $\mathcal{G}_1^{\Omega_{S}}$ & $\mathcal{G}_1^{\Omega_{V}}$ & $\mathcal{G}_2^{A_{+}}$ & $\mathcal{G}_2^{A_{-}}$ \\ \hline		
$\mathcal{M}~(\times 10^{-3})$ & $4.8239$ & $4.8300$ & $4.8306$ & $4.8261$ & $4.8382$ & $4.8344$ & $4.8369$ & $4.8330$ & $4.8333$ & $4.8343$  \\
$\Psi~(\times 10^{1})$ & $5.9541$ & $6.0152$ & $6.1554$ & $6.0376$ & $6.0900$ & $6.0256$ & $5.9666$ & $5.9881$ & $5.9319$ & $6.0644$  \\
		\hline
	\end{tabular}
\end{table}

\begin{figure*}[]
	\centering
	\includegraphics[width=0.95\linewidth]{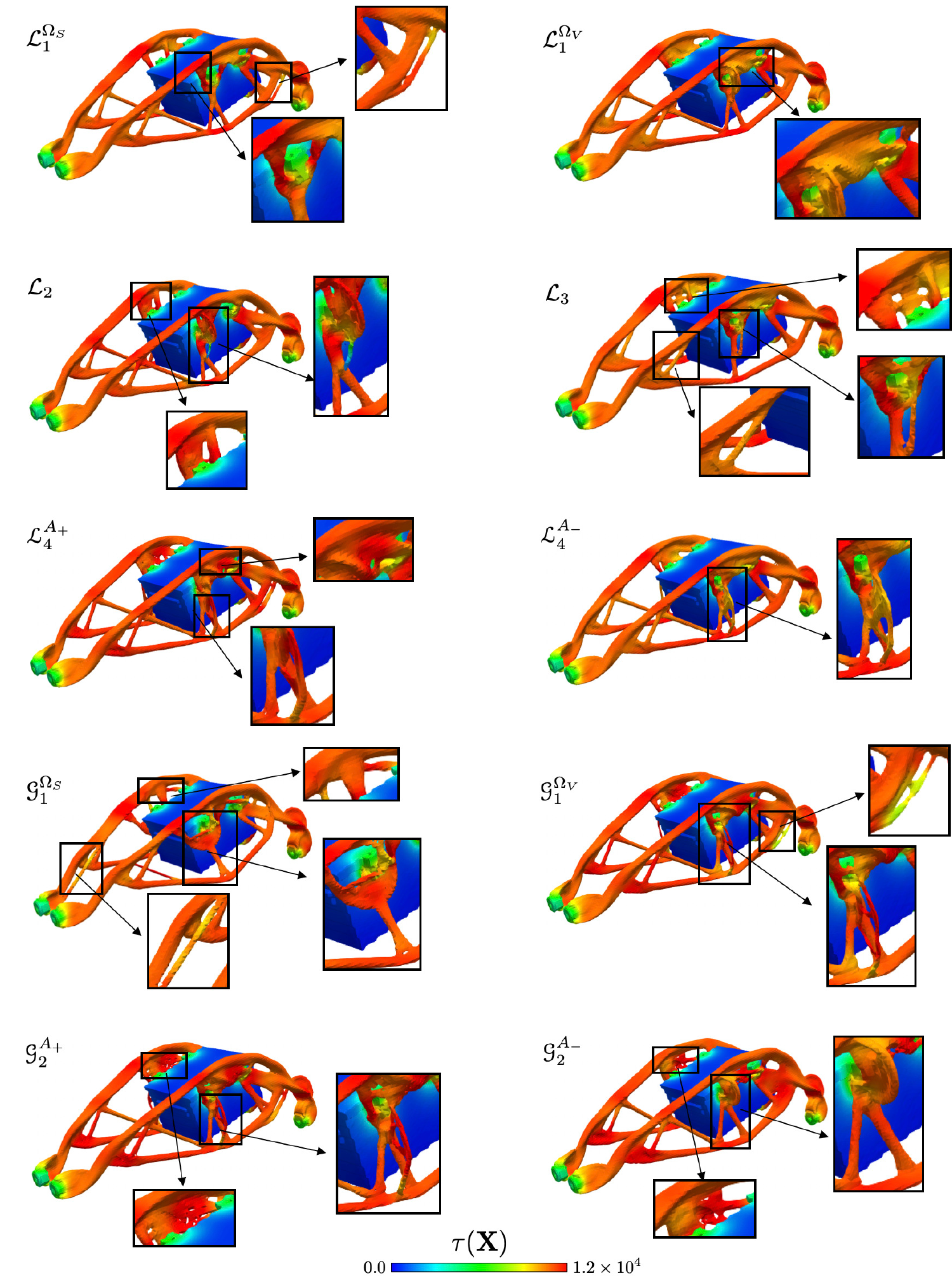}
	\caption{Bracket problem optimized designs using all local and global rules for resolving topology ambiguities.}
	\label{fig:bracketFinalDesigns}
\end{figure*}

The optimized designs with snapshots of regions with noticeable local variations in the final geometries are shown in Fig.~\ref{fig:bracketFinalDesigns}. 
Major differences are observed in the connections between the bolts and design subdomain.
The trusses connecting the main frame of the bracket to these bolts contain between one to multiple bars, and shear-web-like features in some cases.
Thin walls, some with holes, are also observed in the rear region of the design subdomain, close to the back supports.
For all but $\mathcal{L}_1^{\Omega_{V}}$ and $\mathcal{G}_2^{A_{-}}$ rules, additional thinner bars complement the main frame connecting the box and the front supports. 

Overall, although the outer frame of the bracket is similar in all cases, secondary features with various shapes and sizes appear in various regions of the design domain depending on the rule used to resolve $\text{ATs}$. 
A general trend of these features in terms of location cannot be inferred.
Fluctuations in the stress distributions can also be observed. 
Regions of maximum stress are concentrated on thinner features in rules $\mathcal{L}_4^{A_{+}}$, $\mathcal{G}_1^{\Omega_{V}}$, and $\mathcal{G}_2^{A_{+}}$.
Nevertheless, the objective convergence behavior is comparable in all cases, and the differences in performance between results using any of the rules is negligible.

%Figure 

%-------------------------------------------------------------------------------------------------%
\section{Summary and Future Work} \label{sec:concl}
%-------------------------------------------------------------------------------------------------%

The effect of ambiguities in the phase assignment of tetrahedralized intersected hexahedra was analyzed in the context of level set topology optimization. 
To ease the analysis of the underlying issue, these ambiguities were classified as internal ambiguous tetrahedra (i.e., all faces of an ambiguous tetrahedron inscribed within the hexahedron), and boundary ambiguous tetrahedra (i.e., one of the ambiguous tetrahedron faces coinciding with a face of the containing hexahedron).
Inconsistencies encountered for specific intersection patterns were resolved through rules applied locally (i.e., individually for each intersected hexahedron), or globally (i.e., across connected intersected hexahedra).
Using local rules, consistent phase assignment of boundary ambiguous tetrahedra across intersected elements was enforced via the asymptotic decider. In addition, internal ambiguous tetrahedra were resolved using criteria based on predetermined user-defined or geometrical indicators.
Conversely, in the global rules described in this paper, a single criterion is applied to both boundary and internal ambiguous tetrahedra.

The sensitivity of the design process and performance of optimized designs on the aforementioned phase assignment rules were investigated using an explicit level set topology optimization framework.
Although shape and topological changes, such as small dents in the interface or holes in regions with thin features, can appear, they were not observed in the optimized designs.
The minor influence of the ambiguous tetrahedra resolution schemes on the optimized design is likely due to the ability of the optimization process to implicitly account for the phase assignment rule. 
Distinct geometrical configurations developed initially lead to very similar discretized geometries in optimized designs.
Overall, the numerical experiments showed that the optimized designs achieved similar performances for various optimization problem formulations regardless of the rule employed. 
None of the rules examined appeared to be superior to the others, nor any particular rule seems to adversely affect the optimization process.
As long as a rule is used consistently throughout the optimization process, a smooth evolution of the objective can be expected regardless of the rule used.
Nevertheless, we recommend that studies on 3D level set topology using immersed boundary methods with explicit interface representations, such as the $\text{XFEM}$, explicitly state the criteria used for resolving topological ambiguities.
This enhances the reproducibility of results and, thus, an informed analysis on features formed in the optimized designs. 

Additional rules that promote smoothness in the interface while not restricting the evolution of designs, such as curvature measures, should be studied. Furthermore, the numerical efficiency and robustness of the rules provided should be tested for solid-solid domains, and various physical models. 

% -------------------------------------------------------------------
\section*{Compliance with Ethical Standards} \label{sec:RepOfRes}
% --------------------------------------------------------------------

The authors declare that they have no conflict of interest. 
Upon request, the authors will provide the full set of input parameters for each topology optimization problem presented in the paper. 
Any optimization framework with an implementation of the rules described here to determine a phase for ambiguous tetrahedra should be able to reproduce the results shown in the numerical examples section.

%----------------------------%
%-- Acknowledgements --%
%----------------------------%

\section*{Acknowledgments}
\label{acknowledgments}
All authors acknowledge support of the Air Force Office of Scientific Research (grant FA9550-16-1-0169). The first author acknowledges support from NSF (Grant 1463287). The second author acknowledges support from the Defense Advanced Research Projects Agency (DARPA) under the TRADES program (Agreement HR0011-17-2-0022). The opinions and conclusions presented in this paper are those of the authors and do not necessarily reflect the views of the sponsoring organizations.

%----------------------------%
%-----  Bibliography ------%
%----------------------------%

%\section*{References}
%\label{references}

%\biboptions{sort&compress} 
\bibliographystyle{elsarticle-num} 
\bibliography{newRefBiblio}

\end{document}